\documentclass[10pt, conference]{IEEEtran}
\usepackage{t1enc}
\usepackage{times}
\usepackage{amsfonts}

\newcommand{\set}[1]{\left\{ #1\right\}}
\newcommand{\gilt}{:}
\newcommand{\sodass}{\,:\,}
\newcommand{\setGilt}[2]{\left\{ #1\sodass #2\right\}}

\newcommand{\realrange}[2]{\left[#1, #2\right]}

\newcommand{\unitrange}[2]{\realrange{0}{1}}

\newcommand{\llabel}[1]{\label{\labelprefix:#1}}
\newcommand{\labelprefix}{} % later redefined using renewcommand

\newcommand{\discussionsize}{\small}

\newenvironment{code}{\noindent%\sf%
\begin{tabbing}%
\hspace{2em}\=\hspace{2em}\=\hspace{2em}\=\hspace{2em}\=\hspace{2em}\=%
\hspace{2em}\=\hspace{2em}\=\hspace{2em}\=\hspace{2em}\=\hspace{2em}\=%
\kill}{\end{tabbing}}

\newcommand{\labelcommand}{}
\newcommand{\captiontext}{}
\newsavebox{\codeparam}
\newcounter{lineNumber}
\newenvironment{disscodepos}[3]{%
\renewcommand{\labelcommand}{#2}%
\renewcommand{\captiontext}{#3}%
\sbox{\codeparam}{\parbox{\textwidth}{#3}}%
\begin{figure}[#1]\begin{center}\begin{code}\setcounter{lineNumber}{1}}{%
\end{code}\end{center}\caption{\llabel{\labelcommand}\captiontext}\end{figure}}

{\end{disscodepos}}

\newcommand{\Is}       {:=}

\newdimen\endofsize\endofsize=0.5em
\def\endofbeweis{~\quad\hglue\hsize minus\hsize
                 \hbox{\vrule height \endofsize width
\endofsize}\par}
\usepackage{epsfig}
\usepackage{url}
\usepackage{amsmath}
\usepackage{amssymb}
\usepackage{pdflscape}
\usepackage{numprint}
\npdecimalsign{.} % we want . not , in numbers
\usepackage{theorem}
\usepackage{makeidx}

\usepackage[left=2cm,top=1.5cm,right=2cm]{geometry}

\def\MdR{\ensuremath{\mathbb{R}}}

\usepackage{color}

\newcommand{\mytitle}{Think Locally, Act Globally: \\ Perfectly Balanced Graph Partitioning}
\begin{document}
\title{\mytitle}
\author{Peter Sanders, Christian Schulz\\ 
	\textit{Karlsruhe Institute of Technology},
	\textit{Karlsruhe, Germany} \\
	\textit{Email: \{\url{sanders}, \url{christian.schulz}\}\url{@kit.edu}} }
\date{}

\maketitle
\begin{abstract}
We present a novel local improvement scheme for the \textit{perfectly} balanced graph partitioning problem.
This scheme encodes local searches that are not restricted to a balance constraint into a model allowing us to find combinations of these searches maintaining balance by applying a negative cycle detection algorithm. 
We combine this technique with an algorithm to balance unbalanced solutions and integrate it into a parallel multilevel evolutionary algorithm, KaFFPaE, to tackle the problem.
Overall, we obtain a system that is fast on the one hand and on the other hand is able to improve or reproduce most of the best known \textit{perfectly} balanced partitioning results ever reported in the literature. 
\end{abstract}

\thispagestyle{empty}
\section{Introduction}
In computer science, engineering, and related fields \textit{graph partitioning} is a common technique. 
For example, in parallel computing good partitionings of unstructured graphs are very valuable.
In this area, graph partitioning is mostly used to partition the underlying graph model of computation and communication.
Roughly speaking, nodes in this graph represent computation units and edges denote communication. 
This graph needs to be partitioned such that there are few edges between the blocks (pieces). 
In particular, if we want to use $k$ processors we want to partition the graph into $k$ blocks of about equal size. 
In this paper we focus on the \textit{perfectly balanced} version of the problem that constrains the maximum block size to average block size and tries to minimize the total cut size, i.e. the number of edges that run between blocks. 
In practice the perfectly balanced version of this problem is important for small graph models where nodes stand for a large amount of computation. 
If this graph needs to be partitioned into a large number of blocks, e.g. for a large number of processors, then already a small amount of overloaded vertices can yield bad load imbalance. 

During the last years we started to put all aspects of the multi-level graph partitioning (MGP) scheme on trial since we had the impression that certain aspects of the method are not well understood. Our main focus is partition quality rather than partitioning speed.
In our sequential MGP framework KaFFPa (Karlsruhe Fast Flow Partitioner) \cite{kaffpa}, we presented novel local search as well as global search algorithms similar to the strategies used in the multigrid community.
%In our sequential multilevel graph partitioning framework KaFFPa (Karlsruhe Fast Flow Partitioner) \cite{kaffpa}, we presented algorithms based on max-flow min-cut and very localized local search as well as global search algorithms similar to the strategies used in the multigrid community.
In the Walshaw benchmark \cite{soper2004combined}, KaFFPa was beaten mostly for small graphs that combine multilevel partitioning with an evolutionary algorithm. 
We therefore developed an improved evolutionary algorithm, KaFFPaE (KaFFPa Evolutionary) \cite{kaffpaE}, that also employs coarse grained parallelism. 
Both of these algorithms are able to compute partitions of very high quality in a reasonable amount of time when some imbalance $\epsilon > 0$ is allowed. 
However, they are not yet very good for the perfectly balanced case $\epsilon=0$. 
In the perfectly balanced case state-of-the-art local search algorithms are restricted to find nodes to be exchanged between a pair of blocks in order to decrease the cut \textit{and} to maintain perfect balance. 
Hence, we introduce new specialized techniques for the perfectly balanced case in this paper. 
Experiments indicate that the techniques are also useful if some imbalance is allowed. 
From a meta heuristic point of view the proposed algorithms increase the neighborhood of a perfectly balanced solution in which local search is able to find better solutions. Moreover, we provide efficient ways to explore this neighborhood.
As we will see, these algorithms \textit{guarantee} that the output partition is \textit{perfectly balanced} whereas current solvers basically do not guarantee perfect balance.

Although the problem is NP-hard \cite{journals/ipl/BuiJ92} and hard to approximate on general graphs \cite{journals/ipl/BuiJ92} an astonishingly large set of ``easier'' graph algorithms are used to tackle the problem. For example algorithms such as weighted matching, spanning trees, edge coloring, breadth first search, maximum flows, diffusion and strongly connected components.  In this paper this list is further augmented by two well known algorithms: negative cycle detection and shortest path algorithms allowing negative edge weights.

The paper is organized as follows.  
We begin in Section~\ref{s:preliminaries} by introducing basic concepts. 
After shortly presenting Related Work in Section~\ref{s:related}
we describe novel perfectly balanced local search and balancing algorithms in Section~\ref{s:negcyclelocalsearch}. 
Here we start by explaining the very basic idea that allows us to find combinations of simple node movements. 
We then explain directed local searches and extend the basic idea to a complex model containing more node movements. 
This is followed by a description on how these techniques are integrated into KaFFPaE. 
A summary of extensive experiments done to evaluate the performance of our algorithms is presented in Section~\ref{s:experiments}.
\vfill
\section{Preliminaries}\label{s:preliminaries}
\begin{figure}[h!]
\centering
\includegraphics[width=4cm]{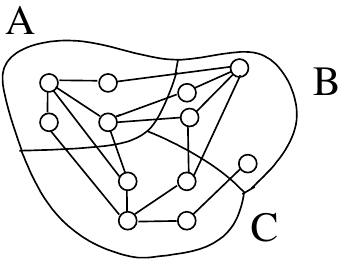}
\includegraphics[width=3.5cm]{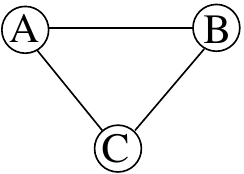}
\caption{A graph that is partitioned into three blocks of size four on the left and its corresponding quotient graph on the right. There is an edge in the quotient graph if there is an edge between the corresponding blocks in the original graph.}
\label{fig:examplequotiengraph}
\end{figure}

%\subsection{Basic concepts}
Consider an undirected graph $G=(V,E,\omega)$ 
with edge weights $\omega: E \to \MdR_{>0}$, $n = |V|$, and $m = |E|$.
We extend $\omega$ to sets, i.e.,
$\omega(E')\Is \sum_{e\in E'}\omega(e)$.
$\Gamma(v)\Is \setGilt{u}{\set{v,u}\in E}$ denotes the neighbors of $v$.
We are looking for \emph{blocks} of nodes $V_1$,\ldots,$V_k$ 
that partition $V$, i.e., $V_1\cup\cdots\cup V_k=V$ and $V_i\cap V_j=\emptyset$
for $i\neq j$. A \emph{balancing constraint} demands that 
$\forall i\in \{1..k\}\gilt |V_i|\leq L_{\max}\Is (1+\epsilon)\lceil |V|/k \rceil $. In the perfectly balanced case the imbalance parameter $\epsilon $ is set to zero.
A graph is perfectly $k$-divisible if $\lceil |V|/k\rceil = |V|/k$.
The objective is to minimize the total \emph{cut} $\sum_{i<j}w(E_{ij})$ where 
$E_{ij}\Is\setGilt{\set{u,v}\in E}{u\in V_i,v\in V_j}$. 
A node $v \in V_i$ that has a neighbor $w \in V_j, i\neq j$, is a boundary node. 
An abstract view of the partitioned graph is the so called \emph{quotient graph}, where nodes represent blocks and edges are induced by connectivity between blocks. An example is shown in Figure~\ref{fig:examplequotiengraph}. Given a partition, the gain of a node $v$ in block $A$ with respect to a block $B$ is defined as $g_{(A,B)}=\omega(\{(v,w)\mid w\in\Gamma(v)\cap B\})- \omega(\{(v,w) \mid w \in \Gamma(v)\cap A)$, i.e. the reduction in the cut when $v$ is moved from block $A$ to block $B$.
By default, our initial inputs will have unit node weights.
However, the algorithms proposed in this paper can be easily extended to weighted nodes.

A successful heuristics for partitioning large graphs is the \emph{multilevel} approach.
Here, the graph is recursively \emph{contracted} to achieve a smaller graph with the same basic structure. 
After applying an \emph{initial partitioning} algorithm to the smallest graph in the hierarchy, the contraction is undone and, at each level, a \emph{local refinement} method is used to improve the partitioning induced by the coarser level.

\section{Related Work}\label{s:related}
There has been a huge amount of research on graph partitioning so that we refer the reader to \cite{GPOverviewBook,fjallstrom1998agp,Walshaw07}.
All general purpose methods that are able to obtain good partitions for large real world graphs are based on the multilevel principle outlined in Section~\ref{s:preliminaries}. 
Well known software packages based on this approach include, Jostle~\cite{Walshaw07}, Metis~\cite{karypis1999pmk}, and Scotch~\cite{Scotch}.  However, for different reasons they are not able guarantee that the produced partition is perfectly balanced.

KaFFPa \cite{kaffpa} is a multi-level graph partitioning algorithm using local improvement algorithms that are based on flows and more localized FM searches. 
KaFFPaE \cite{kaffpaE} is a distributed parallel evolutionary algorithm that uses our multilevel graph partitioning framework KaFFPa \cite{kaffpa} to create individuals and modifies the coarsening phase to provide new effective combine operations. 
It currently holds the best results for many graphs in Walshaw's Benchmark Archive \cite{soper2004combined} when some imbalance is allowed. 
KaPPa \cite{kappa} is a "classical" matching based multi-level graph partitioning algorithm designed for scalable parallel execution. 

DiBaP \cite{meyerhenke2008ndb} is a multi-level graph partitioner where local improvement is based on diffusion. 
It currently holds some of the best results in the perfectly balanced case for large graphs in Walshaw's Benchmark Archive \cite{soper2004combined}.  
Benlic et al. \cite{conf/ictai/BenlicH10,conf/ieeeconftoolsartintell/benlichao2010,journals/cor/BenlicH11} provided multilevel memetic algorithms for perfectly balanced graph partitioning. 
Their approach is able to compute many entries in Walshaw's Benchmark Archive \cite{soper2004combined} for the case $\epsilon=0$. 
However, they are not able to guarantee that the computed partition is perfectly balanced especially for larger values of $k$.
PROBE \cite{journals/tc/ChardaireBM07} is a meta-heuristic which can be viewed as a genetic algorithm without selection. It is restricted to the case $k=2$ and $\epsilon=0$.

\section{Perfectly Balanced Local Search by Negative Cycle Detection}
\label{s:negcyclelocalsearch}
In this section we describe our local search and balancing algorithms for perfectly balanced graph partitioning.
Roughly speaking, all of our algorithms consist of two components. The \textit{first component} are local searches on pairs of blocks that share a non-empty boundary, i.e. all edges in the quotient graph. 
These local searches are not restricted to the balance constraint of the graph partitioning problem and are undone after they have been performed. 
The \textit{second component} uses the information gathered in the first component. 
That means we build a model using the node movements performed in the first step enabling us to find combinations of those node movements that \textit{maintain balance}.

We begin by describing the very basic algorithm and go on by presenting an advanced model which enables us to combine complex local searches. 
This is followed by a description on how local search and balancing algorithms are put together. 
At the end of this section we show how we integrate these algorithms into our evolutionary framework KaFFPaE.

\subsection{Basic Idea\,--\,Using A Negative Cycle Detection Algorithm}
\label{subsec:negcycleeasy}
We are now ready to explain the basic idea for a balanced local search algorithm. 
As we will see, local searches are fairly simple in this case. Before we start we introduce two notations:
a node in the graph $G$ can have two states \textit{marked} and \textit{unmarked}. By default a node is unmarked. 
A node is called \textit{eligible} if it is not adjacent to a previously marked node. 
\vfill

\begin{figure}[t]
\centering
\includegraphics[width=4cm]{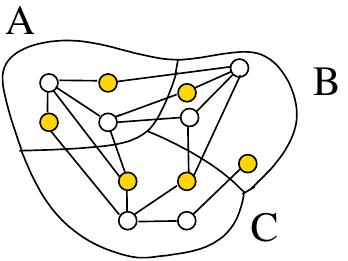}
\includegraphics[width=3.5cm]{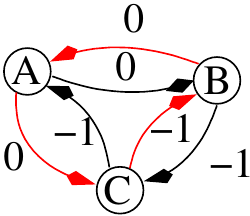}
\includegraphics[width=4.5cm]{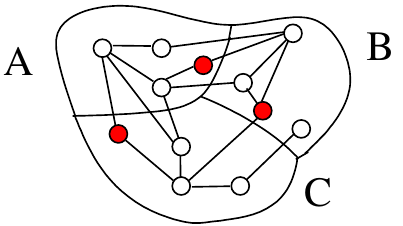}
\caption{On top left an example graph that is partitioned into three parts (A, B and C) of four nodes each. Possible candidates for movement are highlighted. On top right the corresponding model is shown and one negative cycle is highlighted. On the bottom the updated partition after the node movements associated with the cycle are performed is shown. Moved nodes are highlighted. The reduction in the number of edges cut is equal to the weight of the cycle.}
\label{fig:examplepart}
\vspace*{-.5cm}
\end{figure}

We can now build the model of the underlying partition of the graph $G$,
$\mathcal{Q}=(\{1,\cdots,k\},\mathcal{E})$ where $(A,B) \in \mathcal{E}$ if there is an edge in $G$ that runs between the blocks $A$ and $B$. 
We define edge weights $\omega_{\mathcal{Q}}: \mathcal{E} \to \MdR$ in the following way: for each \textit{directed} edge $e = (A, B) \in \mathcal{E}$ in a random order, find a \textit{eligible} boundary node $v$ in block $A$ having maximum gain $g_{\text{max}}(A,B)$, i.e. a node $v$ that maximizes the reduction in cut size when moving it from block $A$ to block $B$. If there is more than one of such nodes we break ties randomly.
The node is then marked.
This is basically all the local search that is done in the basic algorithm. 
The weight of $e$ is then defined as the negative gain value of its associated node $v$, i.e. $\omega_{\mathcal{Q}}(e) := - g_{\text{max}}(A, B)$. 
It is important to notice that a forward edge $(A, B)$ does not have to have the same weight as its backward edge $(B, A)$.
An example partitioned graph with a basic model is shown in Figure~\ref{fig:examplepart}.
Observe that the basic model is basically a directed and weighted version of the quotient graph and that the selected nodes form an independent set.
Note that each cycle in this model defines a set of node movements and furthermore when the associated nodes of a cycle are moved, then each block contains the same number of nodes as before. 
Also the weight of a cycle in the model is equal to the reduction in the cut when the associated node movements are performed.
However, the most important aspect is that a \textit{negative cycle} in the model corresponds to a set of node movements that will decrease the overall cut and maintain the balance of the partition. 
To detect a negative cycle in this model we introduce a node $s$ and connect it to all nodes in $\mathcal{Q}$. 
The weight of the inserted edges is set to zero. 
We can apply a standard shortest path algorithm \cite{golderbergNegCycle} that can handle negative edge weights to detect a negative cycle.
If the model contains a negative cycle we can perform a set of node movements that will not alter block sizes since each block obtains and emits a node.

When starting with an unbalanced partitioning, i.e. a partition of the graph that is not perfectly balanced, or if the graph is not perfectly $k$-divisible, the block weight invariant still holds. 
However, we can add a slight extension to the model. 
That is we connect each node in $\mathcal{Q}$ that corresponds to a block which can take at least one node without becoming overloaded to $s$ by an edge that has weight zero. 
Note that a negative cycle \textit{containing} $s$ may alters some block weights and can lead to a balancing operation reducing the cut.
In fact such a cycle can correspond to a set of node movements starting in an arbitrary block and ending in a block that can take a node without becoming overloaded (basically a path in the quotient graph).
However, if there is no negative cycle in the model we have to think about diversification and balancing strategies which is done in the following sections. 
An interesting observation is that the algorithm can be seen as an extension of the classical FM algorithm that swaps nodes between two adjacent blocks (two at a time) which is basically a negative cycle of length two in our model if the gain of the two node movements is positive.

\subsubsection*{Diversification by  Zero Gain Cycle Moves}
A zero weight cycle in the basic model is associated with a set of node movements that keep the cut unchanged and the block weights constant. 
After such a movement is performed it might be possible to find further negative cycles in the basic model since candidates for movements and gain values may have changed. 
Hence these cycles can be useful to introduce some diversification.

Nonetheless, on general graphs it is NP-complete to decide whether a weighted graph contains a cycle that has weight zero, i.e. the sum of the edge weights of this cycle is zero. 
However, we will see that if a graph does not contain a negative cycle we can decide whether it contains a cycle of weight zero in polynomial time and output one if one exists. 
This can be done by using the following technique.
As soon as the model described above does not contain negative cycles we compute a shortest path tree starting at $s$. 
By doing this we obtain node potentials $\Pi: \{1,\cdots,k\} \to \MdR$, i.e. the shortest path distances from $s$ to all other nodes. 
We then define modified edge weights $\ell_\mathcal{Q}(e=(A,B)) = \omega_\mathcal{Q}(e) + \Pi(A) - \Pi(B)$. 
It is quite easy to see that the weight of a cycle in $\mathcal{Q}$ does not change when we use $\ell_\mathcal{Q}$ instead of $\omega_\mathcal{Q}$.
In particular cycles that have weight zero w.r.t $\omega_\mathcal{Q}$ will have weight zero w.r.t. $\ell_\mathcal{Q}$. Another important observation is that $\ell_\mathcal{Q}$ is a non-negative function. Hence, in order to detect a zero weight cycle we can evict all edges $e$ with $\ell_\mathcal{Q}(e) > 0$ since they cannot be a part of a cycle having weight zero. 
After this is done we compute all strongly connected components of this graph. 
If there is a strongly connected component that contains more than one node then the graph contains a cycle that has weight zero. 
To output one zero weight cycle we pick a random node $N$ out of the components having more than one node. 
Starting at this node we perform a random walk in its component until we find a node that we have already seen $M$ (which is not necessary $N$).
It is then fairly simple to output the respective cycle starting at $M$. Note that if the component contains $j$ nodes than the random walk will stop after at most $j$ iterations. As soon as we have found a cycle of weight zero we can perform the node movements that are associated with the edges of the cycle.

\subsection{Advanced Model}
Our advanced model is strongly coupled with advanced local search algorithms. 
Roughly speaking, each edge in the advanced model is associated with a whole set of node movements which have been found by local searches.
Hence, a negative cycle in this model will correspond to a combination of local searches with positive overall gain that maintain balance or that can improve balance.% if the input partition is not balanced.

Before we build the advanced model we perform a single \textit{directed local search} on each pair of blocks that share a non-empty boundary, i.e. each pair of blocks that are adjacent in the quotient graph. 
A local search on a directed pair of blocks $(A,B)$ is only allowed to move nodes from block $A$ to block $B$. 
The order in which the directed local search between a directed pair of blocks is performed is random. 
That means we pick a random directed adjacent pair of blocks on which local search has not been performed yet and perform local search as described below. 
This is done until local search was done between all directed adjacent pairs of blocks once.

\subsubsection*{Directed Local Search} We now explain how we perform a directed local search between a pair $(A, B)$ of blocks. 
A directed local search between two blocks $A$ and $B$ is very localized akin to the multi-try method used in KaFFPa \cite{kaffpa}. 
However, a directed local search between $A$ and $B$ is restricted to move nodes from block $A$ to block $B$. 
It is similar to the FM-algorithm \cite{fiduccia1982lth}: We start with a \textit{single} random eligible boundary node of block $A$ having maximum gain $g_{\text{max}}(A,B)$ and put this node into a priority queue. 
The priority queue contains nodes of the block $A$ that are valid to move. 
The priority is based on the \emph{gain}, i.e. the decrease in edge cut when the node is moved from block $A$ to block $B$. 
We always move the node that has the highest priority to block $B$.
After a node is moved its eligible neighbors that are in block $A$ are inserted into the priority queue.
We perform at most $\tau$ steps per directed local search, where $\tau$ is a parameter. 
Note that during a directed local search we only move nodes that are not incident to a node moved during a previous directed local search. 
This restriction is necessary to keep the model described below accurate.
Thus we \textit{mark all nodes} touched during a directed local search \textit{after} it was performed which as well implies that each node is moved at most once. 
In addition all moved nodes are \textit{moved back} to their origin, since these movements would make the partition imbalanced. 
We stress that all nodes incident to nodes that have been moved during a directed local search are not \textit{eligible} for any later local search during the construction since this would make the gain values computed imprecise. 
\begin{figure}[t]
\centering
\includegraphics[width=6cm]{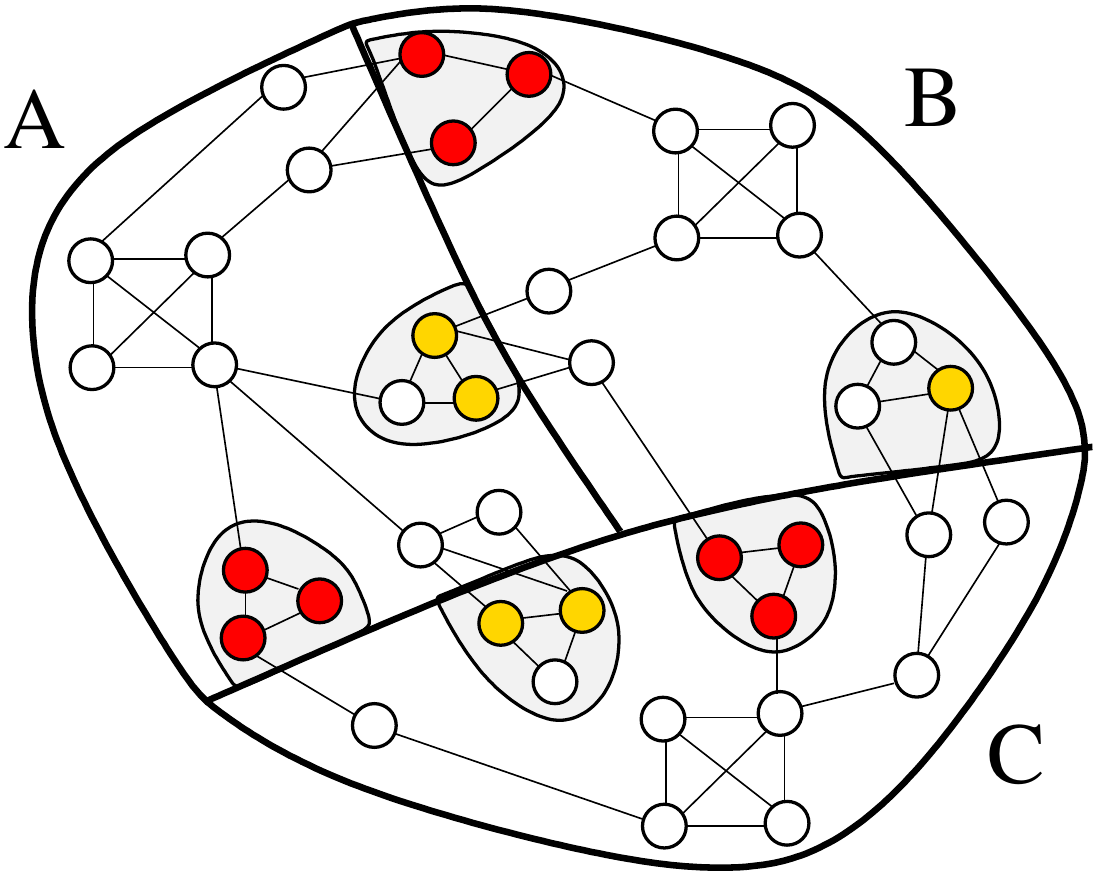} \\
                       \vspace*{.5cm}
\includegraphics[width=6cm]{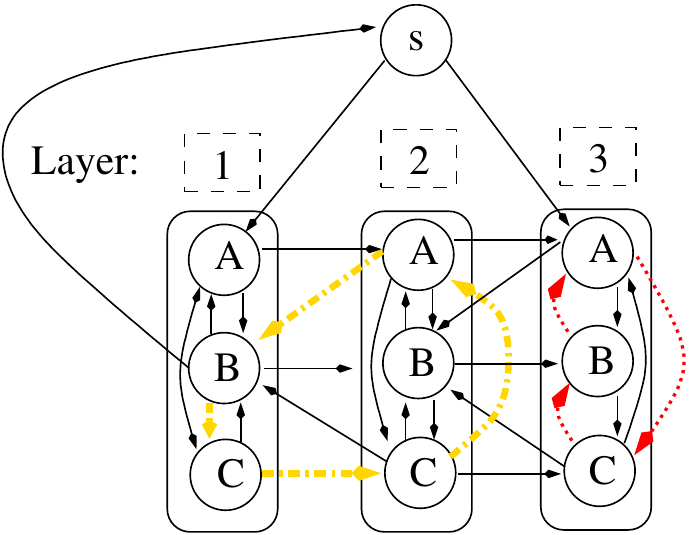}
\caption{On top a graph that is partitioned into three parts ($|A|=14, |B|=12, |C|=14$). Directed local searches on each directed pair of blocks are highlighted ($\tau=3$). The corresponding  advanced model is shown on the bottom. Each layer is a copy of the quotient graph of the partition. Edges within layer $d$ represent node movements consisting of $d$ nodes that have been found previously using directed local search. $s$ is connected to all nodes (most of the edges are not shown), edges back to $s$ are inserted if the corresponding block can take some nodes without becoming overloaded (in this example block $B$), backward edges between layers are inserted if the block can take nodes without becoming overloaded, forward edges between the layers are inserted in any case. Within layer 3 a negative cycle is highlighted (red/dark dashed) which corresponds the movement of the nine red/dark nodes on top.  Another negative cycle is highlighted in yellow/light grey dashed. It corresponds to the movement of the five yellow/light gray nodes on top. The weight of both cycles is -2. After these movements are performed the resulting partition is perfectly balanced.  }
\label{fig:advancedmodelexample}
\vspace*{-.5cm}
\end{figure}
\vspace*{-.25cm}
\subsubsection*{The Model}The advanced model allows us to find combinations of directed local searches such that the balance of the given partition at least maintained.  Specifically a negative cycle in the model represents a set of node movements that maintain or improve balance while the number of edges cut is reduced. Roughly speaking, most of the edges in the advanced model will be associated with a subset of a directed local search that we was computed above.

The local search process described above yields for each pair of blocks $e=(A,B)$ in the quotient graph a sequence of node movements $S_e \in V^{\hat\tau}$ and a sequence of gain values $g_e \in \MdR^{\hat\tau}$. Here $\hat\tau$ is smaller or equal to $\tau$, the maximum number of node movements allowed for a single directed local search.
The $d$'th value in $g_e$ corresponds to the reduction in the cut between the pair of blocks $(A,B)$ when the first $d$ nodes in $S_e$ are moved from their source block $A$ to their target block $B$. 
By construction, a node $v \in V$ can occur  in at most one of the sequences created and in its sequence only once.

Roughly speaking, the \textit{advanced model} consists of $\tau$ layers. 
Essentially each layer is a copy of the quotient graph. 
An edge starting and ending in layer $d$ of this model corresponds to the movement of exactly $d$ nodes. 
The weight of an edge $e=(A, B)$ in layer $d$ of the model is set to the negative value of the $d$'th entry in $g_e$ if $|g_e| \geq d$ otherwise the edge is removed from that layer. 
In other words it encodes the negative value of the gain, when the first $d$ nodes in $S_e$ are moved from block $A$ to block $B$.
Hence, a negative cycle whose nodes are all in layer $d$ will move exactly $d$ nodes between each of the respective block pairs contained in the cycle and results in a overall decrease in the edge cut. 
We add additional edges to the model such that it contains \textit{more moves} in the case of an imbalanced input partition or if the graph is not perfectly $k$-divisible. 
To be more precise in these cases  we want to get rid of the restriction that each block sends and emits the same amount of nodes.
To do so we insert \emph{forward} edges between all consecutive layers, i.e. block $k$ in layer $d$ is connected by an edge of weight zero to block $k$ in layer $d+1$. These edges are not associated with node movements.
Furthermore, we add \emph{backward} edges as follows: for an edge $(A, B)$ in layer $d$ we add an edge with the same weight between block $A$ in layer $d$ and block $B$ in layer $d-\ell$ if block $B$ can take $\ell$ nodes without becoming overloaded. 
The newly inserted edge is associated with the same node movements as the initial edge $(A, B)$ within layer $d$. 
This way we encode movements in the model where a block can emit more nodes then it gets and vice versa without violating the balance constraint.
Additionally we connect each node in layer $d$ back to $s$ if the associated block can take at least $d$ nodes without becoming overloaded. 
As in the basic model this ensures that the model might contain cycles through $s$. 
That means that we also can find cycles corresponding to paths in the quotient graph being associated with node movements that decrease the overall cut. 
Moreover, these moves never increase the imbalance of the input partition.
An example advanced model is shown in Figure~\ref{fig:advancedmodelexample}.
Note that if the input partition is perfectly balanced and the graph is perfectly $k$-divisible then the backward edges, including those back to $s$, are not contained in the model. 
Also note that the zero gain diversification can also be applied in the advanced model.
\subsubsection*{Packing} The algorithm can be further improved by performing/packing multiple directed local searches between each pair of blocks that share a non-empty boundary. 
To be more precise after we have computed node movements on \textit{each} pair of blocks $e=(A,B)$ we start again using the nodes that are still eligible. 
This is done $\mu$ times. We say $\mu$ is the number of packing iterations. 
The model is then slightly modified in the following way: For the creation of edges in the model that correspond to the movement of $d$ nodes from block $A$ to block $B$ we use the directed local search on $e=(A,B)$ from the process above with the best overall gain when moving $d$ nodes from block $A$ to block $B$ (and use this gain value for the computation of the weight of corresponding edges). 
\subsubsection*{Conflicts}
\begin{figure}[t]
\centering
\includegraphics[width=5.5cm]{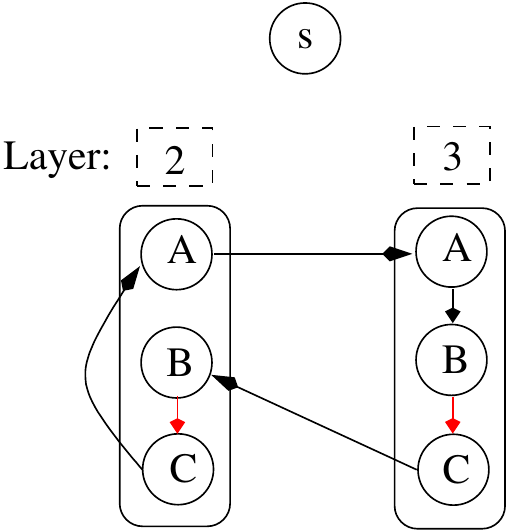}
\caption{The first type of conflict that can occur in the advanced model. In this example two layers of the advanced model corresponding to graph in Figure~\ref{fig:advancedmodelexample} are shown. Only the edges of a conflicted cycle are drawn. The problem of the cycle are highlighted edges running in layer two and three from the nodes representing block B to the nodes representing block C. They are associated with node movements where subsets are equal. The drawn cycle does not correspond to a simple cycle in the quotient graph.}
\label{fig:cyclenotsimple}
\end{figure}

The advanced model is a bit problematic since it contains two types of conflicts due to the edges that run between the layers. 
\textit{First} the model can contain cycles that do not correspond to a \emph{simple} cycle in the quotient graph. 
Such a cycle is problematic because it contains the \emph{same} edge $e=(A, B)$ in the quotient graph multiple times. 
An example is given in Figure~\ref{fig:cyclenotsimple}.
Let us assume that one associated edge runs in layer $d$ and one in layer $\ell$ with $\ell < d$. 
The associated node movements cannot be performed fully since the edges correspond to subsets of the \emph{same} directed local search. 
This is due to the fact that the edge in layer $\ell$ corresponds to the movement of the first $\ell$ nodes in $S_e$.
These movements are a subset of the node movements associated with the edge in layer $d$, which corresponds to movement of the first $d$ nodes in $S_e$. 
In other words, when we want to move the nodes associated with the edge in layer $d$ then they are already in block $B$ if the node movements of the edge in layer $\ell$ have been performed before and vice versa depending on the order of execution. 
That means that for at least one of those two edges its weight does not correspond to the reduction in the cut of the underlying node movements.
Hence, the weight of the cycle does not reflect the reduction in the number of edges cut. 
\begin{figure}[t]
\centering
\includegraphics[width=5.5cm]{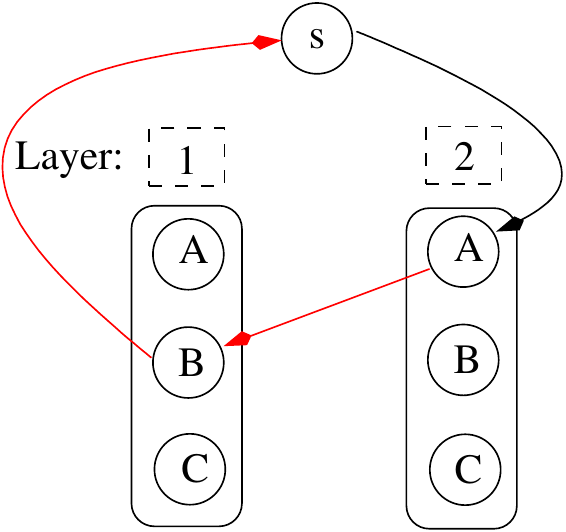}
\caption{The second type of conflict that can occur in the advanced model. In this example two layers of the advanced model from Figure~\ref{fig:advancedmodelexample} are drawn. Only the edges of a conflicted cycle are shown. The edge in layer one from the node representing block $B$ back to $s$ was created because block $B$ can take one node without becoming overloaded ($|B|=12, |V|/k=13$). For the same reason there is the edge between the layers from the node representing block A in layer two to the node representing block B in layer one. In the model there is no edge from block B in layer two back to $s$ since block two can only take one node without becoming overloaded. However, when performing the associated node movements block B receives two vertices from block A and is overloaded afterwards.}
\label{fig:overloadconflict}
\vspace*{-.5cm}
\end{figure}

\textit{Secondly} since we have both, edges between the layers \textit{and} the edges back to $s$, a cycle in the model can lead to node movements that \emph{overload} a block. A example is given in Figure~\ref{fig:overloadconflict}.  
A conflict can only occur if we have edges running between the layers which is only the case when we start with an unbalanced input partition or if the graph is not perfectly $k$-divisible.
Our experiments indicate that conflicts do not occur very often. 
Furthermore, a conflicted cycle is easily detected. We can simply check if the cycle in the model is a simple cycle in the quotient graph or if one block would get overloaded when performing the node movements of that cycle.
If our algorithm returns a cycle that contains a conflict we \textit{remove a random edge} of the cycle in the model and start the negative cycle detection strategy again.
Note that if we remove all edges in the model that run between the layers then the model is conflict-free but encodes less possible combinations of node movements.

\subsubsection*{Balancing}
As we will see in Section~\ref{sec:integrationintoKaFFPaE}, to create perfectly balanced partitions we start our algorithm with an $\epsilon$-balanced partition, i.e. a partition where larger imbalance is allowed.
Hence, to achieve perfect balance we have to think about balancing strategies.
A balancing step will only be applied if the model does not contain a negative cycle (see next section for more details).
Hence, we can modify the advanced model such that we can find a set of node movements that will decrease the total number of overloaded nodes by at least one and  minimizes the increase in the number of edges cut. 
Specifically, we introduce a second node $t$. 
Now instead of connecting $s$ to all vertices, we connect it only to nodes representing overloaded blocks, i.e. $|V_i| > \lceil |V|/k\rceil$. 
Additionally, we connect a node in layer $\ell$ to $t$ if the associated block can take at least $\ell$ nodes without becoming overloaded. 
Since the underlying model does not contain negative cycles we can apply a shortest path algorithm to find a shortest path from $s$ to $t$. 
We use a variant of the algorithm of Bellman and Ford since edge weights might still be negative (for more details see Section \ref{sec:implementation}).
It is now easy to see that a shortest path in this model yields a set of node movements with the smallest loss in number of cut edges and that the total number of overloaded nodes decreases by at least one. 
If $\tau$ is set to one we call this algorithm basic balancing otherwise advanced balancing.

However, we have to make sure that there is at least one $s$-$t$ path in the model. Let us assume for now that the graph is connected.
If the graph is connected then the directed version of the quotient graph is strongly connected. 
Hence a $s$-$t$ path exists in the model if we are able to perform local search between  \textit{all} pairs of blocks that share a non-empty boundary.
Because a directed local search can only start from an eligible node we might not be able to perform directed local search between all adjacent pairs of blocks, e.g. if there is no eligible node between a pair of blocks left.
We try to \textit{ensure} that there is at least one $s$-$t$ path in the model by doing the following.
Roughly speaking we try to integrate a $s$-$t$ path in the model by changing the order in which directed local searches are performed.
First we perform a breadth first search (BFS) in the quotient graph which is initialized with all nodes that correspond to overloaded blocks in a random order. 
We then pick a random node in the quotient graph that corresponds to a block $A$ that can take nodes without becoming overloaded.
Using the BFS-forest we find a path $\mathcal{P}=B \to \cdots \to A$ from an overloaded block $B$ to $A$. 
We now first perform directed local search on all consecutive pairs of blocks in $\mathcal{P}$. Here we use $\tau=1$ for the number of node movements to minimize the number of non-eligible nodes. 
If this was successful, i.e.  we have been able to move one node between all directed pairs of blocks in that path, we perform directed local searches as before on \textit{all} pairs of blocks that share a non-empty boundary. 
Otherwise we undo the searches done (every node is eligible again) and start with the next random block that can take a node without becoming overloaded.

\begin{figure}[t]
\centering
\includegraphics[width=6cm]{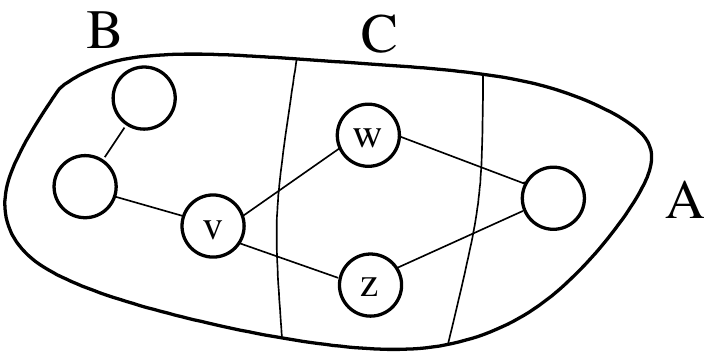} \\
                       \vspace*{.5cm}
\includegraphics[width=3.75cm]{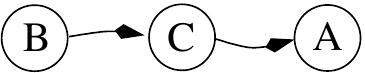} 
\caption{On top an example graph that is partitioned into three parts and on bottom a BFS-tree in the quotient graph starting in overloaded block $B$. It is not possible to integrate this path into the model since after directed local search is done on the pair $(B,C)$, $v$ will be marked and hence there is no eligible node left for the local search on the pair $(C,A)$. A similar argument holds if local search is done on the pair $(C,A)$ first.}
\label{fig:examplerebfail}
\vspace*{-.5cm}
\end{figure}

In some rare cases the algorithm fails to find such a path, i.e. each time we look at a path we have one directed pair of blocks where no eligible node is left. An example is shown in Figure~\ref{fig:examplerebfail}.
In this case we apply a fallback balance routine that guarantees to reduce the total number of overloaded nodes by one if the input graph is connected. 
Given the BFS-forest  of the quotient graph from above we look at all paths in it from an overloaded block to a block that can take a node without becoming overloaded. At this point there are at most order of $k$ such paths in our BFS-forest.
Specifically for a path $\mathcal{P}=Z \to Y \to X \to \cdots \to A$ we select a node having maximum gain $g_{Z,Y}$ in $Z$ and move it to $Y$. 
We then look at $Y$ and do same with respect to $X$ and so on until we move a node to block $A$. 
Note that this time we can assure to find nodes because after a node has been moved it is not blocked for later movements. 
After the operations have been performed they are undone and we continue with the next path. In the end we use the movements of the path that resulted in the smallest number of edges cut.

If the graph contains more than one connected component then the algorithms described above may not work. For example if there is a non-perfectly balanced block in the input partition that is the union of some of the graphs connected components. 
More precisely, when we want to integrate a path into the model we detect at some point that there is no path in quotient graph that contains this block and that can yield a balance improvement, e.g. if the block corresponds to a singleton in the quotient graph.
To reduce the total number of overloaded nodes by one we do the following:
If the block is overloaded we move a random node from this block to an underloaded block; otherwise we move a random node from an overloaded block to this block. 

Note that the advanced balancing model can contain conflicts, too.  
This is again because of the edges that run between the layers. 
We handle potential conflicts in paths analogously to the conflicts in the advanced model case. 

\subsubsection*{Putting Things Together}
In practice we start our algorithms with an unbalanced input partition (see Section~\ref{sec:integrationintoKaFFPaE} for more details). We define two algorithms basic and advanced depending on the models used. Both the basic and the advanced algorithm operate in rounds. 
In each round we iterate the negative cycle based local search algorithm until there are no negative cycles in the corresponding model (basic or advanced).  
After each negative cycle local search step we try to find zero weight cycles in the model to introduce some diversification. In Section~\ref{sec:walshawbenchmarkimproving} we also use a variant of the basic algorithm that does not use zero weight cycle diversification.
Since we have random tie breaking at multiple places we iterate this part of the algorithm. 
If we do not succeed to find an improved cut using these two operations for $\lambda$ iterations we perform a single balancing step if the partition is still unbalanced otherwise we stop. 
The parameter $\lambda$ basically controls how fast the unbalanced input partition is transformed into a partition that satisfies the balance constraint.
After the balancing operation, the total number of overloaded nodes is reduced by at least one depending on the balancing model. In the basic algorithm we use the basic balancing model ($\tau=1$) and in the advanced algorithm we use the advanced balancing model.
Since the balance operation can introduce new negative cycles in the model we start the next round. We call the refinement techniques introduced in this paper Karlsruhe Balanced Refinement (KaBaR).

\subsection{Integration into KaFFPaE}
\label{sec:integrationintoKaFFPaE}
We now describe how we integrate our new algorithms into our distributed evolutionary algorithm KaFFPaE \cite{kaffpaE}. 
An evolutionary algorithm starts with a population of individuals (in our case partitions of the graph) and evolves the population into different populations over several rounds. 
In each round, the evolutionary algorithm uses a selection rule based on the fitness of the individuals (in our case the edge cut) of the population to select good individuals and combine them to obtain improved offspring. 
Roughly speaking, KaFFPaE uses KaFFPa to create individuals and modifies the coarsening phase to provide new effective combine operations. 

It is well known that allowing larger imbalance is useful to create good partitions \cite{walshaw2000mpm,simon1997good}. Hence, we adopt this idea. To obtain perfectly balanced partitions we modify the create and combine operations as follows: each time we perform such an operation, we randomly choose an imbalance parameter $\epsilon' \in [0.005,\hat\epsilon]$ where $\hat\epsilon$ is an upper bound for the allowed imbalance (a tuning parameter). This imbalance is then used  to perform the operation, i.e. after the operation is performed, the offspring/partition has blocks with size at most $(1+\epsilon')\lceil |V|/k \rceil$. 
Giving a larger imbalance to the operation yields smaller cuts and makes local search more effective since the combine and create operations use the multilevel graph partitioner KaFFPa. 
After the respective operation is performed we apply our advanced balancing and advanced negative cycle local search (including zero weight cycle diversification and the packing approach) to obtain a partition of the graph that is \textit{perfectly} balanced. 
This individual is the final offspring created by the performed create or combine operation and inserted into the population using the techniques of KaFFPaE \cite{kaffpaE}. 
Note that \textit{at all times} each individual in the population of the evolutionary algorithm is \textit{perfectly balanced}.
Also note that allowing larger imbalance enables us to use previous developed techniques that otherwise would not be applicable, e.g. max-flow min-cut based local search methods.
We call the overall algorithm Karlsruhe Balanced Partitioner Evolutionary (KaBaPE).
As experiments will show in Section~\ref{s:experiments} the new kind of local search is also helpful if some imbalance is allowed. 
When we use KaBaPE to create $\epsilon$-balanced partitions we choose $\epsilon' \in [\epsilon+0.005,\epsilon+\hat\epsilon]$ for the combine and create operations. 
The created individual is then transformed into a partition where each block has size at most $(1+\epsilon)\lceil |V|/k \rceil$ using our balancing and negative cycle local search strategies. 
\vfill
\subsection{Miscellanea}
We also tried to integrate the negative cycle detection strategies into the multi-level scheme of KaFFPa.
However, experiments did not indicate large improvements and furthermore the runtime increased drastically. 
This is due to the fact that the size of the model of the negative cycle detection strategies depends heavily on the sum of the \textit{weights} of the nodes moved (the number of layers in the model is the maximum of the sum of the weights moved between a pair of blocks during construction of the directed local searches). 
Also recall that a multi-level graph partitioning algorithm creates a sequence of smaller graphs, e.g. by computing matchings and contracting matched edges. 
This kind of compression is not helpful for our model in the perfectly balanced refinement scheme.
\section{Experiments}
\label{s:experiments}
\subsubsection*{Implementation}
\label{sec:implementation}
We have implemented the algorithm described above using C++. We implemented negative cycle detection with subtree disassembly and distance updates as described in \cite{golderbergNegCycle}. 
Overall,
our program (including KaFFPa(E)) consists of about 23\,000 lines of code. The implementation of the perfectly balanced local search algorithms has about 3\,400 lines of code.
\subsubsection*{System}
Experiments have been done on two machines. 
Machine A has four Quad-core Opteron 8350 (2.0GHz), 64GB RAM, running Ubuntu 10.04. 
Machine B is a cluster with 200 nodes where each node is equipped with two Quad-core Intel Xeon processors (X5355) which run at a clock speed of 2.667 GHz. 
Each node has 2x4 MB of level 2 cache each and run Suse Linux Enterprise 11 SP 1.  
All nodes are attached to an InfiniBand 4X DDR interconnect which is characterized by its very low latency of below 2 microseconds and a point to point bandwidth between two nodes of more than 1300 MB/s.
All programs were compiled using GCC Version 4.7 and optimization level~3 using OpenMPI 1.5.5.
\subsubsection*{Parameters}
\label{sec:parametertuning}
After an extensive evaluation of the parameters we fixed the number of packing iterations to $\mu=20$ (larger values of $\mu$, e.g. iterating until no boundary node is eligible did not yield further improvements).
The maximum number of node movements per directed local search is set to $\tau=15$ for $k\leq 8$ and to $\tau = 7$ for $k > 8$ since this turned out to be a good tradeoff between quality and runtime. The number of unsuccessful iterations until we perform a balancing step $\lambda$ is set to three. 
When using KaBaPE to create perfectly balanced or $\epsilon$-balanced partitions we choose random values around the parameters above for each create or combine operation. 
To be more precise, each time we perform a create or combine operation we pick a random number of node movements per directed local search $\tau \in [1, 30]$, a random number of packing iterations $\mu \in [1, 20]$ and $\lambda \in [1,10]$ and use these parameters for the balancing and negative cycle detection strategies.

\subsection{Walshaw Benchmark}
\label{sec:walshawbenchmark}
In this section we apply our techniques to all graphs in Chris Walshaw's benchmark archive \cite{soper2004combined}.
This archive is a collection of real-world instances for the graph partitioning problem. The rules used there imply that the running time is not an issue, but one wants to achieve minimal cut values for $k \in \{2,4,8,16,32,64\}$ and balance parameters $\epsilon \in \{0,0.01,0.03,0.05\}$. 
It is the most used graph partitioning benchmark in the literature. 
Most of the graphs of the benchmark come from finite-element applications; however, there are also some graphs from VLSI design and a road network. 
In KaFFPa and KaFFPaE we focused on partitions of graphs where some imbalance, i.e. $\epsilon \in \{0.01,0.03,0.05\}$, is allowed since the techniques used therein are not made for the case $\epsilon=0$.

\subsubsection*{Improving Existing Partitions}
\label{sec:walshawbenchmarkimproving}
When we started to look at perfectly balanced partitioning we counted the number of perfectly balanced partitions in the benchmark archive that contain nodes having positive gain, i.e. nodes that could reduce the cut when being moved to a different block. 
Astonishingly, we found that 55\% of the perfectly balanced partitions in the archive contain nodes with positive gain (some of them have up to 1400 of such nodes). 
These nodes usually cannot be moved by simple local search due to the balance constraint. 
Therefore, we now use the existing perfectly balanced partitions in the benchmark archive and use them as input to our local search algorithms KaBaR. This experiment has been performed on machine A and for all configurations of the algorithm we used $\lambda=20$ for the number of unsuccessful tries.
Table~\ref{tab:existingimprovment} shows the relative number of partitions that have been improved by different algorithm configurations and $k$ (in total there are 34 graphs per number of blocks $k$). 
%15:48:08 algorithm_evaluation_walshaw $ ./count_improvements.sh 
%===================== algorithm: basic
%2 0
%4 6
%8 13
%16 22
%32 26
%64 28
%overall 95
%===================== algorithm: basicpluszero
%2 0
%4 8
%8 17
%16 23
%32 26
%64 28
%overall 102
%===================== algorithm: advwithoutpacking
%2 0
%4 14
%8 22
%16 24
%32 30
%64 27
%overall 117
%===================== algorithm: advwithpacking
%2 0
%4 15
%8 25
%16 27
%32 31
%64 30 
%overall 128

\begin{table}[t]
\centering
 \begin{tabular}{|l|r|r|r|r|}
 %\hline
 $k$    & Basic & +ZeroGain & Advanced &+Packing\\ 
 \hline                                 
 \hline                                 
 2      & 0\%   & 0\%       & 0\%      &\textbf{0\%} \\ 
 %\hline                                 
 4      & 18\%  & 24\%     & 41\%     & \textbf{44\%} \\ 
 %\hline                                  
 8      & 38\%  & 50\%     & 64\%     & \textbf{74\%} \\ 
 %\hline                                 
 16     & 64\%  & 68\%     & 71\%     & \textbf{79\%} \\ 
 %\hline                                 
 32     & 76\%  & 76\%     & 88\%     & \textbf{91\%}\\ 
 %\hline                                 
 64     & 82\%  & 82\%      & 79\%     & \textbf{88\%}\\ 
 \hline                                 
 \hline                                 
 sum    & 47\%  & 50\%     & 57\%     & \textbf{63\%}\\ 
 %\hline
\end{tabular}
        \caption{Relative number of improved instances in the Walshaw Benchmark. Configurations: Basic (Most Basic Negative Cycle Improvement), +ZeroGain (As Before Plus Zero Weight Cycle Diversification), Advanced (Advanced Model, Directed Local Searches and Zero Weight Cycle Diversification), +Packing (As Before Plus Packing Enabled)}
        \label{tab:existingimprovment}
        \vspace*{-.5cm}
\end{table}

It is somewhat surprising that already the most basic variant of the algorithm, i.e. negative cycle detection without the zero weight cycle diversification mechanism, can improve 47\% of the existing entries.
All of the algorithms increase the number of improved partitions with increased number of blocks $k$ (except the two advanced algorithms). 
Less surprisingly but still interesting to see is that the more advanced the local searches and models are becoming the more partitions can be improved. 
Note that it took roughly two hours to compute 63\% entries, i.e. 128 partitions, having a smaller cut than reported in the archive using one core of machine A when applying the advanced algorithm with packing enabled (the most expensive configuration of the algorithm). 
This is \textit{very affordable} considering the fact that some of the previous approaches, such as Soper et. al. \cite{soper2004combined}, have taken many days to compute \textit{one}  entry to the benchmark tables. 
Of course in practice we want to find high quality partitions without using input partitions generated by other algorithms. We therefore compute partitions from scratch in the next section.
\subsubsection*{Computing Partitions from Scratch}
We now compute perfectly balanced partitions from scratch, i.e. we do not take existing partitions as input to our algorithm. 
To do so we use machine B and run KaBaPE with a time limit $t_k=225\cdot k$ seconds using 32 cores (four nodes of the cluster) per graph and $k>2$. 
On the eight largest graphs of the archive we gave KaBaPE a time limit of $\hat t_k = 4\cdot t_k$ seconds per graph and $k>2$. 
For $k=2$ we gave KaBaPE one hour of time and 32 cores. 
$\hat\epsilon$ was set to 4\% for the small graphs and to 3\% for the eight largest graph in the archive. 
We summarize the results in Table~\ref{tab:wsscratch} and report the complete list of results obtained in the Appendix.
Currently we are able to improve or reproduce 86\% of the entries reported in this benchmark.
In the bipartition case we mostly reproduce the entries reported in the benchmark (instead of improving). 
This is not surprising since the models presented in this paper can contain only trivial cycles of length two in this case and our previous algorithms have shown the same behaviour for larger imbalance values (\cite{kappa, kaspar, kaffpa, kaffpaE}). 
Also recently it has been shown by Delling et. al \cite{delling2012better} that some of the perfectly balanced bipartitions reported there are optimal.
We also applied our algorithm for larger imbalances, i.e. 1\%, 3\% and 5\%, in the Walshaw Benchmark. 
For the case $\epsilon=1\%$ we run our algorithm KaBaPE on all instances using the same parameters $\hat\epsilon$ and $t_k$ as in the perfectly balanced case. Here we are able to improve or reproduce the cut in 160 out of 204 cases. A table reporting detailed results can be found in the Appendix.
Afterwards we performed additional partitioning trials on all instances where our systems (including \cite{kappa}. \cite{kaspar}. \cite{kaffpa}, \cite{kaffpaE}) currently \textit{not} have been able to reproduce or improve the entry reported there using different parameters and different machines. 
Doing so our systems  now improved or reproduced 
98\%, 99\%, 99\%, 99\% of the entries reported there for the cases $\epsilon=0, 1\%, 3\%, 5\%$ respectively. These numbers include the entries where we used the current record as an input to our algorithms. 
They contribute roughly 4\%, 7\%, 11\%, 9\% for the cases $\epsilon=0,1\%,3\%,5\%$ respectively. 
%An overview of the improvements  that have been computed from scratch with respect to the current records is shown in Figure~\ref{fig:improvmentscratchperfectly}. 
\begin{table}[h!]
\centering
\scriptsize
\vspace*{-.5cm}
\begin{tabular}{|l|r|r|r|r|r|r|r|}
%\hline
$k$           & \textbf{2}  & \textbf{4}& \textbf{8} & \textbf{16} & \textbf{32} & \textbf{64} & $\sum$\\
\hline
\hline
<      & 4 & 19&24&25&30&29& 64\%\\ 
                \hline
$\leq$ & 29 &31&27&27&31&30 & 86\%\\
                \hline

 %\begin{tabular}{|c|c|c|}
%%\hline
%$k$&$<$ & $\leq$\\
                %\hline
                %\hline
                %2             & 4     & 29 \\
                %%\hline
                %4             & 19    & 31 \\
                %%\hline
                %8             & 24    & 27 \\
                %%\hline
                %16            & 25    & 27 \\
                %%\hline
                %32            & 30    & 31 \\
                %%\hline
                %64            & 29    & 30 \\
                %\hline
                %\hline
                %overall       & 64 \% & 86 \%\\
                %\hline
 \end{tabular}
\caption{Number of improvements computed from scratch $\epsilon=0$.}
\label{tab:wsscratch}
\vspace*{-1cm}
\end{table}
%\vspace*{-.5cm}
%\begin{wraptable}{r}{9cm}
%\centering
 %\begin{tabular}{||c|c|c||}
%%\hline
%$k$&$<$ & $\leq$\\
                %\hline
                %\hline
                %2             & 2     & 26 \\
                %\hline
                %4             & 16    & 27 \\
                %\hline
                %8             & 22    & 25 \\
                %\hline
                %16            & 21    & 23 \\
                %\hline
                %32            & 28    & 29 \\
                %\hline
                %64            & 29    & 30 \\
                %\hline
                %\hline
                %overall       & 57 \% & 77 \%\\
                %%\hline
 %\end{tabular}
%\caption{Number of improvements computed from scratch.}
%\end{wraptable}

%\begin{figure}[h]
%\centering
%\includegraphics[width=0.45\textwidth]{tables/plot.pdf}
%\caption{Improvement relative to previous records.}
%\label{fig:improvmentscratchperfectly}
%\end{figure}
%\flush
\pagebreak
\subsection{Costs for Perfect Balance}
It is hard to perform a meaningful comparison to other partitioners since publicly available tools such as Scotch \cite{Scotch}, Jostle~\cite{Walshaw07} and Metis~\cite{karypis1999pmk} are either not able to take the desired balance as an input parameter or are not able to guarantee perfect balance. 
This is a major problem for the comparison with these tools since allowing larger imbalances, i.e. $\epsilon=3\%$, decreases the number of edges cut significantly \cite{simon1997good}. 
Hence, we have a look at the number of edges cut by our algorithm when perfect balance is enforced, i.e. the increase in the number of edges cut when we seek a perfectly balanced partition.  
To do so we use machine B and KaFFPaStrong to create partitions having an imbalance of $\epsilon=1\%$ and then create perfectly balanced partitions using our advanced negative cycle model and advanced balancing. 
KaFFPaStrong is designed to achieve very good partition quality and is the strongest configuration of our multi-level graph partitioner KaFFPa.
For each instance (graph, $k$) we repeat the experiment ten times using different random seeds. We then compare the final cuts of the perfectly balanced partitions to the number of edges cut before the balancing and negative cycle search started, i.e. when $\epsilon=1\%$ imbalance is allowed.
The instances used for this experiment are the same as in KaFFPa \cite{kaffpa} and are available for download at the 10th DIMACS Implementation Challenge~\cite{benchmarksfornetworksanalysis,dimacschallengegraphpartandcluster}. The main properties of these graphs are summarized in the Appendix. 
Table~\ref{tab:costsforperfectbalance} summarizes the results of the experiment, detailed results are reported in the Appendix. 
On average the number of edges cut increased by roughly 6\% when enforcing perfect balance and the runtime of the negative cycle local search and balancing strategies is comparable with the average runtime of KaFFPaStrong. 

\begin{table}[h!]
\centering
\scriptsize
\begin{tabular}{|l|r|r|r|r|r|r|}
%\hline
$k$           & \textbf{2}  & \textbf{4}& \textbf{8} & \textbf{16} & \textbf{32} & \textbf{64}\\
\hline
\hline
Rel. Cut      & 9\% & 7\%&5\%&6\%&4\%&3\%\\ 
                \hline
Rel. time $t$ & 12\% &56\%&99\%&107\%&134\%&163\%\\
                \hline
%2& 642/589& 3.35/2.99 \\
%4& 1462/1365&7.15/4.57 \\
%8& 2564/2450& 14.31/7.20\\
%16&4212/4001& 23.83/11.48\\
%32&6606/6375& 42.43/18.16\\
%64&10353/10092&81.56/30.97\\
\end{tabular}        
        \caption{Costs for Perfect Balance, Rel. to KaFFPaStrong when $\epsilon=1\%$ imbalance is allowed. Rel. Cut reports the average increase in the cut after the 1\% partitions have been balanced and Rel. time reports the average time used by KaBaR relative to the runtime of KaFFPaStrong.}
        \label{tab:costsforperfectbalance}
        \vspace*{-.5cm}
\end{table}

\section{Conclusion and Future Work}
In this paper we have presented novel algorithms to tackle the \textit{perfectly balanced} graph partitioning problem.
These algorithms are able to combine local searches by a model in which a cycle corresponds to a set of node movements in the original partitioned graph that roughly speaking do not alter the balance of the partition. Here we demonstrated that a negative cycle detection algorithm is very well suited to find cycles in our model that are associated with node movements decreasing the overall cut. Experiments indicate that previous algorithms have not been able to find such rather complex movements.
An integration into our parallel multi-level evolutionary algorithm is able to improve or reproduce \textit{most} of the entries reported in the Walshaw Benchmark in a reasonable amount of time. Additionally the algorithm is also useful if some imbalance is allowed. 
In contrast to previous algorithms such as Scotch \cite{Scotch}, Jostle~\cite{Walshaw07} and Metis~\cite{karypis1999pmk}, our algorithms are able to \textit{guarantee} that the output partition is feasible. 

An open question is whether it is possible to define a \textit{conflict-free} model that encodes the same kind of node movements as our advanced model.
In future work, it could be interesting to see if one can integrate other type of local searches from KaFFPa~\cite{kaffpa} into our models.  
The packing algorithm can be improved such that instead of simply picking the best directed local search moving $d$ nodes, it could find the best combination of the computed directed local searches to move $d$ nodes. 
It will be also interesting to see whether our techniques are applicable to other problems where local search is restricted by constraints. 
For example this kind of local search could be very interesting in the area of multi-constraint graph partitioning or in the area of hypergraph partitioning.

\subsubsection*{Acknowledgements}
Financial support by the Deutsche Forschungsgemeinschaft (DFG) is gratefully acknowledged (DFG grant SA 933/10-1).
\bibliographystyle{plain}
\bibliography{quellen}
\begin{appendix}
 \begin{table}[h]
\footnotesize
\begin{center}
\begin{tabular}{|l|r|r|}
\hline
\multicolumn{3}{|l|}{Test instances} \\
\hline
graph & $n$ & $m$ \\
\hline
rgg17	 & $2^{17}$ 	& 1\,457\,506\\
rgg18 	 & $2^{18}$  	& 3\,094\,566\\
Delaunay17	 & $2^{17}$	& 786\,352\\
Delaunay18 	 & $2^{18}$ 	& 1\,572\,792\\
\hline
bcsstk29 	&	13\,992 &	605\,496\\
4elt 		&	15\,606 & 	91\,756\\
fesphere  	& 	16\,386 & 	98\,304\\
cti  		&	16\,840 & 	96\,464\\	
memplus 	&	17\,758 &	108\,384\\
cs4 	 	&       33\,499 & 	87\,716\\
pwt 		&	36\,519 & 	289\,588\\
bcsstk32 	&	44\,609 &	1\,970\,092\\	
body 		& 45\,087 	& 327\,468\\
t60k 		& 60\,005 	& 178\,880\\
wing  		& 62\,032 	& 243\,088\\
finan512 	& 74\,752 	& 522\,240\\	
rotor & 99\,617 & 1\,324\,862 \\
\hline
bel & 463\,514 & 1\,183\,764\\
nld & 893\,041 & 2\,279\,080\\
\hline
af\_shell9 	& 504\,855 & 17\,084\,020\\
\hline
\end{tabular}
\end{center}
\caption{Instances used in the cost for perfect balance experiment.}
\end{table}
\thispagestyle{empty}
\begin{table*}[H]
\begin{center}
\tiny
\vspace*{-.75cm}
%\hspace*{.5cm}
\begin{tabular}{l|r||rrr|rrr|rrr|rrr|}
\hline
                  &   $k$ & \multicolumn{3}{c|}{KaFFPa $\epsilon=1\%$}                                                                       & \multicolumn{3}{c|}{KaFFPa $\epsilon=1\%$ + Balancing to $\epsilon=0$}            \\ 
\hline
                  &    & Best.                           & Avg.                                  & total time $t$.                          & Best.                          & Avg.   & total time $t$.      \\
\hline
4elt              & 2  & 138                             & 146                                   & 0.30                        & 142                            & 149    & 0.32    \\
4elt              & 4  & 323                             & 364                                   & 0.57                        & 327                            & 370    & 0.75    \\
4elt              & 8  & 540                             & 587                                   & 1.02                        & 550                            & 593    & 1.49    \\
4elt              & 16 & 957                             & 992                                   & 1.74                        & 963                            & 1001   & 2.33    \\
4elt              & 32 & 1625                            & 1659                                  & 2.42                        & 1619                           & 1659   & 3.12    \\
4elt              & 64 & 2672                            & 2707                                  & 3.47                        & 2670                           & 2700   & 4.30    \\
\hline
af\_shell9        & 2  & 8825                            & 8825                                  & 101.65                      & 8881                           & 8881   & 108.10  \\
af\_shell9        & 4  & 20925                           & 21638                                 & 128.21                      & 21069                          & 21729  & 146.27  \\
af\_shell9        & 8  & 41800                           & 43102                                 & 120.02                      & 42244                          & 43478  & 173.46  \\
af\_shell9        & 16 & 76700                           & 79090                                 & 162.48                      & 76948                          & 79445  & 362.62  \\
af\_shell9        & 32 & 126200                          & 128752                                & 211.97                      & 126744                         & 129228 & 680.10  \\
af\_shell9        & 64 & 195475                          & 198480                                & 269.85                      & 196239                         & 199209 & 1753.50 \\
\hline
bcsstk29          & 2  & 2818                            & 2838                                  & 1.45                        & 2846                           & 2860   & 1.59    \\
bcsstk29          & 4  & 8404                            & 9210                                  & 2.15                        & 8449                           & 9247   & 2.81    \\
bcsstk29          & 8  & 14323                           & 15155                                 & 3.24                        & 14470                          & 15081  & 8.12    \\
bcsstk29          & 16 & 23367                           & 23981                                 & 4.82                        & 23141                          & 24060  & 10.95   \\
bcsstk29          & 32 & 37032                           & 37781                                 & 6.00                        & 36921                          & 37565  & 10.81   \\
bcsstk29          & 64 & 58968                           & 60248                                 & 9.89                        & 57838                          & 58972  & 30.10   \\
\hline
bcsstk32          & 2  & 4667                            & 5234                                  & 6.87                        & 4842                           & 5390   & 8.19    \\
bcsstk32          & 4  & 9251                            & 10214                                 & 8.65                        & 9324                           & 10428  & 15.99   \\
bcsstk32          & 8  & 20419                           & 24738                                 & 12.23                       & 20718                          & 25043  & 46.33   \\
bcsstk32          & 16 & 36930                           & 38923                                 & 13.69                       & 37239                          & 39264  & 58.04   \\
bcsstk32          & 32 & 63843                           & 65520                                 & 17.60                       & 63975                          & 65295  & 112.79  \\
bcsstk32          & 64 & 98093                           & 100016                                & 27.43                       & 97676                          & 98682  & 153.62  \\
\hline
bel               & 2  & 78                              & 82                                    & 10.63                       & 81                             & 104    & 10.90   \\
bel               & 4  & 190                             & 200                                   & 16.73                       & 219                            & 263    & 30.93   \\
bel               & 8  & 376                             & 403                                   & 18.62                       & 414                            & 465    & 37.47   \\
bel               & 16 & 667                             & 705                                   & 37.79                       & 771                            & 844    & 85.20   \\
bel               & 32 & 1179                            & 1218                                  & 120.74                      & 1266                           & 1361   & 207.99  \\
bel               & 64 & 1803                            & 1883                                  & 239.71                      & 1909                           & 2063   & 422.93  \\
\hline
cs4               & 2  & 375                             & 387                                   & 0.55                        & 377                            & 390    & 0.57    \\
cs4               & 4  & 962                             & 989                                   & 1.25                        & 969                            & 990    & 1.52    \\
cs4               & 8  & 1484                            & 1514                                  & 3.12                        & 1488                           & 1517   & 4.11    \\
cs4               & 16 & 2163                            & 2196                                  & 5.72                        & 2163                           & 2198   & 7.32    \\
cs4               & 32 & 3024                            & 3058                                  & 9.87                        & 3025                           & 3060   & 12.13   \\
cs4               & 64 & 4192                            & 4224                                  & 20.92                       & 4191                           & 4225   & 23.70   \\
\hline
cti               & 2  & 318                             & 363                                   & 0.46                        & 334                            & 369    & 0.53    \\
cti               & 4  & 952                             & 1019                                  & 0.99                        & 955                            & 1023   & 1.52    \\
cti               & 8  & 1789                            & 1826                                  & 1.92                        & 1815                           & 1850   & 3.33    \\
cti               & 16 & 2932                            & 2991                                  & 3.87                        & 2954                           & 3022   & 5.92    \\
cti               & 32 & 4305                            & 4384                                  & 7.02                        & 4335                           & 4406   & 10.06   \\
cti               & 64 & 6085                            & 6185                                  & 11.76                       & 6074                           & 6172   & 14.48   \\
\hline
delaunay17     & 2  & 613                             & 622                                   & 4.81                        & 613                            & 636    & 5.38    \\
delaunay17     & 4  & 1218                            & 1267                                  & 6.44                        & 1248                           & 1294   & 9.82    \\
delaunay17     & 8  & 2263                            & 2312                                  & 10.48                       & 2290                           & 2349   & 22.30   \\
delaunay17     & 16 & 3643                            & 3715                                  & 16.86                       & 3686                           & 3772   & 34.00   \\
delaunay17     & 32 & 5825                            & 5970                                  & 23.77                       & 5904                           & 6062   & 66.36   \\
delaunay17     & 64 & 8767                            & 8883                                  & 28.46                       & 8870                           & 8999   & 95.89   \\
\hline
delaunay18     & 2  & 861                             & 875                                   & 12.05                       & 879                            & 901    & 12.88   \\
delaunay18     & 4  & 1726                            & 1751                                  & 15.85                       & 1743                           & 1799   & 24.88   \\
delaunay18     & 8  & 3185                            & 3283                                  & 24.03                       & 3234                           & 3350   & 61.95   \\
delaunay18     & 16 & 5147                            & 5252                                  & 33.73                       & 5286                           & 5368   & 90.50   \\
delaunay18     & 32 & 8211                            & 8360                                  & 47.31                       & 8341                           & 8526   & 183.42  \\
delaunay18     & 64 & 12225                           & 12497                                 & 55.19                       & 12456                          & 12722  & 312.68  \\
\hline
body          & 2  & 262                             & 275                                   & 1.25                        & 262                            & 276    & 1.23    \\
body          & 4  & 617                             & 733                                   & 1.42                        & 626                            & 766    & 3.67    \\
body          & 8  & 1059                            & 1156                                  & 2.68                        & 1114                           & 1224   & 3.98    \\
body          & 16 & 1861                            & 1966                                  & 3.88                        & 1972                           & 2101   & 6.49    \\
body          & 32 & 2991                            & 3142                                  & 5.12                        & 3087                           & 3325   & 11.59   \\
body          & 64 & 5193                            & 5356                                  & 8.37                        & 5249                           & 5476   & 21.20   \\
\hline
pwt           & 2  & 340                             & 363                                   & 0.89                        & 340                            & 365    & 1.00    \\
pwt           & 4  & 705                             & 885                                   & 1.24                        & 713                            & 891    & 1.49    \\
pwt           & 8  & 1448                            & 1535                                  & 1.69                        & 1463                           & 1550   & 2.60    \\
pwt           & 16 & 2813                            & 2867                                  & 2.12                        & 2847                           & 2888   & 3.77    \\
pwt           & 32 & 5581                            & 5720                                  & 3.59                        & 5634                           & 5734   & 7.88    \\
pwt           & 64 & 8371                            & 8474                                  & 12.27                       & 8373                           & 8480   & 19.57   \\
\hline
rotor         & 2  & 2033                            & 2156                                  & 7.12                        & 2098                           & 2222   & 9.60    \\
rotor         & 4  & 7722                            & 8206                                  & 10.85                       & 7838                           & 8314   & 28.21   \\
rotor         & 8  & 13285                           & 13755                                 & 15.77                       & 13521                          & 13941  & 59.46   \\
rotor         & 16 & 21044                           & 22254                                 & 27.81                       & 21243                          & 22468  & 123.37  \\
rotor         & 32 & 33053                           & 33862                                 & 51.35                       & 33385                          & 34028  & 238.95  \\
rotor         & 64 & 48059                           & 48766                                 & 90.82                       & 48339                          & 48994  & 439.99  \\
\hline
fesphere        & 2  & 386                             & 387                                   & 0.39                        & 386                            & 387    & 0.45    \\
fesphere        & 4  & 785                             & 799                                   & 0.83                        & 781                            & 798    & 1.08    \\
fesphere        & 8  & 1207                            & 1231                                  & 1.98                        & 1204                           & 1230   & 2.46    \\
fesphere        & 16 & 1743                            & 1838                                  & 3.19                        & 1744                           & 1836   & 3.93    \\
fesphere        & 32 & 2564                            & 2630                                  & 4.80                        & 2564                           & 2628   & 5.60    \\
fesphere        & 64 & 3775                            & 3828                                  & 7.13                        & 3767                           & 3820   & 7.83    \\
\hline
finan512          & 2  & 162                             & 162                                   & 3.05                        & 162                            & 162    & 2.90    \\
finan512          & 4  & 324                             & 348                                   & 3.11                        & 324                            & 348    & 3.24    \\
finan512          & 8  & 648                             & 754                                   & 3.35                        & 648                            & 754    & 3.50    \\
finan512          & 16 & 1296                            & 1304                                  & 3.62                        & 1296                           & 1304   & 3.83    \\
finan512          & 32 & 2592                            & 2592                                  & 4.93                        & 2592                           & 2592   & 5.73    \\
finan512          & 64 & 10897                           & 11051                                 & 12.73                       & 10953                          & 11118  & 24.31   \\
\hline
memplus           & 2  & 5822                            & 6263                                  & 1.26                        & 5779                           & 6233   & 1.92    \\
memplus           & 4  & 10088                           & 10341                                 & 2.46                        & 9949                           & 10163  & 4.38    \\
memplus           & 8  & 12342                           & 12483                                 & 6.87                        & 12098                          & 12299  & 12.68   \\
memplus           & 16 & 13963                           & 14264                                 & 23.66                       & 13838                          & 14183  & 38.10   \\
memplus           & 32 & 15601                           & 15841                                 & 65.23                       & 15527                          & 15756  & 88.71   \\
memplus           & 64 & 18028                           & 18390                                 & 133.69                      & 17685                          & 18102  & 165.68  \\
\hline
nld               & 2  & 59                              & 65                                    & 23.00                       & 64                             & 120    & 26.14   \\
nld               & 4  & 106                             & 125                                   & 32.84                       & 154                            & 220    & 65.98   \\
nld               & 8  & 249                             & 284                                   & 37.49                       & 325                            & 397    & 115.38  \\
nld               & 16 & 530                             & 572                                   & 58.50                       & 672                            & 813    & 186.11  \\
nld               & 32 & 908                             & 967                                   & 87.30                       & 1177                           & 1276   & 323.82  \\
nld               & 64 & 1536                            & 1651                                  & 139.03                      & 1840                           & 2040   & 591.02  \\
\hline
rgg17 & 2  & 517                             & 540                                   & 6.10                        & 517                            & 682    & 7.36    \\
rgg17 & 4  & 1088                            & 1178                                  & 8.65                        & 1229                           & 1381   & 15.45   \\
rgg17 & 8  & 1933                            & 2113                                  & 12.23                       & 2086                           & 2420   & 41.43   \\
rgg17 & 16 & 3357                            & 3509                                  & 17.30                       & 3804                           & 4146   & 76.91   \\
rgg17 & 32 & 5527                            & 5761                                  & 23.72                       & 6147                           & 6382   & 126.59  \\
rgg17 & 64 & 8954                            & 9306                                  & 31.40                       & 9591                           & 10188  & 257.89  \\
\hline
rgg18 & 2  & 820                             & 851                                   & 18.93                       & 867                            & 1151   & 22.15   \\
rgg18 & 4  & 1703                            & 1752                                  & 25.36                       & 1889                           & 2156   & 49.78   \\
rgg18 & 8  & 3170                            & 3609                                  & 31.06                       & 3484                           & 4156   & 116.56  \\
rgg18 & 16 & 5482                            & 5672                                  & 41.26                       & 6322                           & 6696   & 172.96  \\
rgg18 & 32 & 8850                            & 9412                                  & 47.44                       & 9948                           & 10610  & 314.07  \\
rgg18 & 64 & 13876                           & 14553                                 & 56.68                       & 15059                          & 16093  & 581.57  \\
\hline
t60k              & 2  & 76                              & 79                                    & 1.09                        & 80                             & 82     & 1.18    \\
t60k              & 4  & 208                             & 224                                   & 1.58                        & 210                            & 229    & 1.92    \\
t60k              & 8  & 463                             & 473                                   & 2.24                        & 469                            & 479    & 3.44    \\
t60k              & 16 & 839                             & 857                                   & 3.86                        & 841                            & 862    & 5.51    \\
t60k              & 32 & 1390                            & 1413                                  & 6.02                        & 1401                           & 1420   & 10.41   \\
t60k              & 64 & 2193                            & 2221                                  & 9.65                        & 2203                           & 2224   & 16.05   \\
\hline
wing              & 2  & 811                             & 842                                   & 1.98                        & 811                            & 842    & 2.06    \\
wing              & 4  & 1668                            & 1710                                  & 4.01                        & 1679                           & 1721   & 6.24    \\
wing              & 8  & 2526                            & 2597                                  & 10.00                       & 2536                           & 2608   & 14.79   \\
wing              & 16 & 3914                            & 3992                                  & 17.70                       & 3933                           & 4010   & 24.27   \\
wing              & 32 & 5787                            & 5851                                  & 34.58                       & 5824                           & 5879   & 53.79   \\
wing              & 64 & 7875                            & 7941                                  & 103.40                      & 7896                           & 7976   & 136.66  \\
\hline
\end{tabular}
\caption{Detailed results of the cost for perfect balance experiment.}

\thispagestyle{empty}
\end{center}
\end{table*}

\thispagestyle{empty}
\vfill
\pagebreak

\pagebreak
\section{Detailed Walshaw Benchmark Results}
\label{sec:walshawbenchmarktable}
%\nopagebreak
\begin{landscape}
%\nopagebreak
%\vspace*{-1cm}
\begin{table}[h!]
\scriptsize
\begin{center}
\normalsize
%\vspace*{-2cm}
\begin{tabular}{|l||r|r||r|r||r|r||r|r||r|r||r|r|}\hline
\scriptsize
Graph/$k$  & \multicolumn{2}{|c|}{2} & \multicolumn{2}{|c|}{4} & \multicolumn{2}{|c|}{8} & \multicolumn{2}{|c|}{16} & \multicolumn{2}{|c|}{32} & \multicolumn{2}{|c|}{64}\\

        \hline 
         add20       & \numprint{627}            & \textbf{\numprint{596}}   & \numprint{1181}           & *\textbf{\numprint{1159}}  & +\textbf{\numprint{1693}} & \numprint{1696}            & +\textbf{\numprint{2062}} & \numprint{2063}            & \textbf{\numprint{2421}}   & \numprint{2687}            & \textbf{\numprint{2987}}   & \numprint{3108} \\
 data        & \textbf{\numprint{189}}   & \numprint{189}            & \textbf{\numprint{382}}   & \numprint{382}            & \textbf{\numprint{671}}   & \numprint{679}             & \textbf{\numprint{1133}}  & \numprint{1135}            & \textbf{\numprint{1832}}   & \numprint{1858}            & \textbf{\numprint{2884}}   & \numprint{2885} \\
 3elt        & \textbf{\numprint{90}}    & \numprint{90}             & \textbf{\numprint{201}}   & \numprint{201}            & \textbf{\numprint{345}}   & \numprint{348}             & \textbf{\numprint{573}}   & \numprint{581}             & \textbf{\numprint{965}}    & \numprint{967}             & \textbf{\numprint{1543}}   & \numprint{1553} \\
 uk          & \textbf{\numprint{19}}    & \numprint{19}             & \textbf{\numprint{41}}    & \numprint{41}             & \textbf{\numprint{84}}    & \numprint{89}              & \textbf{\numprint{146}}   & \numprint{150}             & \textbf{\numprint{254}}    & \numprint{280}             & \textbf{\numprint{408}}    & \numprint{417} \\
 add32       & \textbf{\numprint{11}}    & \numprint{11}             & \textbf{\numprint{34}}    & \numprint{34}             & \textbf{\numprint{68}}    & \numprint{75}              & \numprint{126}            & \textbf{\numprint{121}}    & +\textbf{\numprint{217}}   & \numprint{230}             & \numprint{487}             & @\textbf{\numprint{486}} \\
 bcsstk33    & \textbf{\numprint{10171}} & \numprint{10171}          & \textbf{\numprint{21718}} & \numprint{21719}          & \textbf{\numprint{34437}} & \numprint{34579}           & \textbf{\numprint{54911}} & \numprint{55136}           & \textbf{\numprint{77670}}  & \numprint{78054}           & +\textbf{\numprint{108091}}& \numprint{108467} \\
 whitaker3   & \textbf{\numprint{127}}   & \numprint{127}            & \textbf{\numprint{381}}   & \numprint{382}            & \textbf{\numprint{656}}   & \numprint{657}             & \textbf{\numprint{1091}}  & \numprint{1093}            & \textbf{\numprint{1668}}   & \numprint{1697}            & \textbf{\numprint{2514}}   & \numprint{2532} \\
 crack       & \textbf{\numprint{184}}   & \numprint{184}            & \textbf{\numprint{366}}   & \numprint{366}            & \textbf{\numprint{679}}   & \numprint{679}             & \textbf{\numprint{1089}}  & \numprint{1108}            & \textbf{\numprint{1692}}   & \numprint{1728}            & \textbf{\numprint{2551}}   & \numprint{2602} \\
 wing\_nodal & \textbf{\numprint{1707}}  & \numprint{1707}           & \textbf{\numprint{3577}}  & \numprint{3577}           & +\textbf{\numprint{5440}}  & \numprint{5443}            & \textbf{\numprint{8344}}  & \numprint{8422}            & \textbf{\numprint{11913}}  & \numprint{12080}           & \textbf{\numprint{16046}}  & \numprint{16183} \\
 fe\_4elt2   & \textbf{\numprint{130}}   & \numprint{130}            & \textbf{\numprint{349}}   & \numprint{349}            & \textbf{\numprint{607}}   & \numprint{608}             & \textbf{\numprint{1007}}  & \numprint{1009}            & \textbf{\numprint{1614}}   & \numprint{1628}            & \textbf{\numprint{2488}}   & \numprint{2519} \\
 vibrobox    & \numprint{11538}          & \textbf{\numprint{10343}} & \textbf{\numprint{18978}} & \numprint{19098}          & \textbf{\numprint{24680}} & \numprint{24715}           & \numprint{33094}          & *\textbf{\numprint{32197}} & \textbf{\numprint{41504}}  & \numprint{42187}           & \textbf{\numprint{48350}}  & \numprint{50154} \\
 bcsstk29    & \textbf{\numprint{2843}}  & \numprint{2843}           & \textbf{\numprint{8035}}  & \numprint{8159}           & \textbf{\numprint{14232}} & \numprint{14322}           & \textbf{\numprint{22326}} & \numprint{22606}           & \textbf{\numprint{34970}}  & \numprint{35422}           & \textbf{\numprint{56186}}  & \numprint{57435} \\
 4elt        & \textbf{\numprint{139}}   & \numprint{139}            & \textbf{\numprint{326}}   & \numprint{326}            & \textbf{\numprint{545}}   & \numprint{545}             & \textbf{\numprint{939}}   & \numprint{939}             & \textbf{\numprint{1556}}   & \numprint{1562}            & \textbf{\numprint{2587}}   & \numprint{2636} \\
 fe\_sphere  & \textbf{\numprint{386}}   & \numprint{386}            & \textbf{\numprint{768}}   & \numprint{770}            & \textbf{\numprint{1156}}  & \numprint{1165}            & \textbf{\numprint{1714}}  & \numprint{1732}            & \textbf{\numprint{2499}}   & \numprint{2542}            & \textbf{\numprint{3571}}   & \numprint{3625} \\
 cti         & \textbf{\numprint{334}}   & \numprint{334}            & \textbf{\numprint{954}}   & \numprint{954}            & \textbf{\numprint{1788}}  & \numprint{1812}            & \textbf{\numprint{2839}}  & \numprint{2909}            & \textbf{\numprint{4092}}   & \numprint{4129}            & \textbf{\numprint{5762}}   & \numprint{5955} \\
 memplus     & +\textbf{\numprint{5513}} & \numprint{5513}           & +\textbf{\numprint{9554}} & \numprint{9643}           & \numprint{12026}          & *\textbf{\numprint{11845}} & \textbf{\numprint{13159}} & \numprint{13516}           & \textbf{\numprint{14579}}  & \numprint{14634}           & \textbf{\numprint{16277}}  & \numprint{17376} \\
 cs4         & \textbf{\numprint{369}}   & \numprint{370}            & \textbf{\numprint{934}}   & \numprint{934}            & \textbf{\numprint{1440}}  & \numprint{1451}            & \textbf{\numprint{2096}}  & \numprint{2105}            & \textbf{\numprint{2931}}   & \numprint{2938}            & \textbf{\numprint{4027}}   & \numprint{4196} \\
 bcsstk30    & \textbf{\numprint{6394}}  & \numprint{6394}           & +\textbf{\numprint{16653}}& \numprint{16652}          & \textbf{\numprint{34858}} & \numprint{34898}           & \textbf{\numprint{70572}} & \numprint{70681}           & \textbf{\numprint{114990}} & \numprint{119164}          & \textbf{\numprint{173246}} & \numprint{179277} \\
 bcsstk31    & \numprint{2767}           & \textbf{\numprint{2762}}  & \textbf{\numprint{7351}}  & \numprint{7352}           & +\textbf{\numprint{13296}}& \numprint{13361}  & \textbf{\numprint{23876}} & \numprint{24551}           & \numprint{37955}           & *\textbf{\numprint{37652}} & \textbf{\numprint{58258}}  & \numprint{60724} \\
 fe\_pwt     & \textbf{\numprint{340}}   & \numprint{340}            & \textbf{\numprint{705}}   & \numprint{707}            & \textbf{\numprint{1447}}  & \numprint{1450}            & \textbf{\numprint{2831}}  & \numprint{2838}            & \textbf{\numprint{5575}}   & \numprint{5758}            & \textbf{\numprint{8242}}   & \numprint{8495} \\
 bcsstk32    & \textbf{\numprint{4667}}  & \numprint{4667}           & \textbf{\numprint{9314}}  & \numprint{9318}           & \textbf{\numprint{20476}} & \numprint{21099}           & \textbf{\numprint{36251}} & \numprint{36399}           & \textbf{\numprint{60539}}  & \numprint{63856}           & \textbf{\numprint{92284}}  & \numprint{98859} \\
 fe\_body    & \textbf{\numprint{262}}   & \numprint{262}            & \textbf{\numprint{599}}   & \numprint{621}            & \textbf{\numprint{1046}}  & \numprint{1048}            & \textbf{\numprint{1789}}  & \numprint{2057}            & \textbf{\numprint{2942}}   & \numprint{3371}            & \textbf{\numprint{4863}}   & \numprint{5460} \\
 t60k        & \textbf{\numprint{79}}    & \numprint{79}             & \textbf{\numprint{209}}   & \numprint{213}            & \textbf{\numprint{456}}   & \numprint{473}             & \textbf{\numprint{813}}   & \numprint{866}             & \textbf{\numprint{1324}}   & \numprint{1440}            & \textbf{\numprint{2077}}   & \numprint{2233} \\
 wing        & \textbf{\numprint{789}}   & \numprint{790}            & \textbf{\numprint{1627}}  & \numprint{1664}           & \textbf{\numprint{2504}}  & \numprint{2517}            & \textbf{\numprint{3877}}  & \numprint{3890}            & +\textbf{\numprint{5594}}  & \numprint{5626}            & \numprint{7683}            & *\textbf{\numprint{7656}} \\
 brack2      & \textbf{\numprint{731}}   & \numprint{731}            & \textbf{\numprint{3084}}  & \numprint{3084}           & \textbf{\numprint{7142}}  & \numprint{7269}            & +\textbf{\numprint{11644}}& \numprint{11649}           & \textbf{\numprint{17593}}  & \numprint{18229}           & \textbf{\numprint{26135}}  & \numprint{27178} \\
 finan512    & \textbf{\numprint{162}}   & \numprint{162}            & \textbf{\numprint{324}}   & \numprint{324}            & \textbf{\numprint{648}}   & \numprint{648}             & \textbf{\numprint{1296}}  & \numprint{1296}            & \textbf{\numprint{2592}}   & \numprint{2592}            & \textbf{\numprint{10560}}  & \numprint{10560} \\
 fe\_tooth   & \textbf{\numprint{3816}}  & \numprint{3817}           & \textbf{\numprint{6889}}  & \numprint{6917}           & \textbf{\numprint{11418}} & \numprint{11475}           & +\textbf{\numprint{17395}}& \numprint{17396}           & \textbf{\numprint{25130}}  & \numprint{26346}           & \textbf{\numprint{35042}}  & \numprint{35980} \\
 fe\_rotor   & \textbf{\numprint{2098}}  & \numprint{2098}           & \textbf{\numprint{7222}}  & \numprint{7480}           & \numprint{12905}          & *\textbf{\numprint{12864}} & \numprint{20714}          & *\textbf{\numprint{20438}} & \textbf{\numprint{31810}}  & \numprint{32783}           & \textbf{\numprint{47273}}  & \numprint{49381} \\
 598a        & \textbf{\numprint{2398}}  & \numprint{2398}           & \textbf{\numprint{8002}}  & \numprint{8016}           & \numprint{16069}          & *\textbf{\numprint{15924}} & \textbf{\numprint{25753}} & \numprint{26427}           & \textbf{\numprint{38911}}  & \numprint{41538}           & \textbf{\numprint{57418}}  & \numprint{59708} \\
 fe\_ocean   & \textbf{\numprint{464}}   & \numprint{464}            & \textbf{\numprint{1884}}  & \numprint{1886}           & +\textbf{\numprint{4200}}  &\numprint{4200}   & \numprint{8003}           & *\textbf{\numprint{7771}}  & \textbf{\numprint{12806}}  & \numprint{12811}           & \textbf{\numprint{20294}}  & \numprint{22301} \\
 144         & \textbf{\numprint{6486}}  & \numprint{6486}           & \textbf{\numprint{15731}} & \numprint{15804}          & \textbf{\numprint{25288}} & \numprint{26175}           & \textbf{\numprint{38279}} & \numprint{39568}           & \textbf{\numprint{56260}}  & \numprint{58462}           & \textbf{\numprint{79376}}  & \numprint{81973} \\
 wave        & \textbf{\numprint{8677}}  & \numprint{8680}           & \textbf{\numprint{17242}} & \numprint{17475}          & \textbf{\numprint{29229}} & \numprint{30511}           & \textbf{\numprint{42856}} & \numprint{44711}           & \textbf{\numprint{61478}}  & \numprint{65772}           & \textbf{\numprint{85051}}  & \numprint{88986} \\
 m14b        & \textbf{\numprint{3836}}  & \numprint{3836}           & \textbf{\numprint{13074}} & \numprint{13391}          & \textbf{\numprint{25841}} & \numprint{27066}           & \textbf{\numprint{42173}} & \numprint{44541}           & \textbf{\numprint{66328}}  & \numprint{68027}           & \textbf{\numprint{97900}}  & \numprint{101551} \\
 auto        & \numprint{10134}          & \textbf{\numprint{10117}} & \textbf{\numprint{27390}} & \numprint{28048}          & \textbf{\numprint{46041}} & \numprint{48901}           & \textbf{\numprint{77486}} & \numprint{81500}           & \textbf{\numprint{121959}} & \numprint{125477}          & \textbf{\numprint{174553}} & \numprint{176435} \\

        \hline
        \end{tabular}
        \end{center} \caption{Computing partitions from scratch $\epsilon = 0$\%. In each $k$-column the results computed by KaBaPE are on the left and the current Walshaw cuts are presented on the right side. Entries marked with a * can be improved by our refinement techniques when using the current record as input. Entries marked with a + have been obtained using different parameters and seeds. @ indicates that we hold this record.}

        \end{table}
        \end{landscape}
        %\vfill
        %\vspace*{5cm}

\pagebreak
\section{Detailed Walshaw Benchmark Results}
\label{sec:walshawbenchmarktable}
%\nopagebreak
\begin{landscape}
%\nopagebreak
%\vspace*{-1cm}
\begin{table}[h!]
\scriptsize
\begin{center}
\normalsize
%\vspace*{-2cm}
\begin{tabular}{|l||r|r||r|r||r|r||r|r||r|r||r|r|}\hline
\scriptsize
Graph/$k$  & \multicolumn{2}{|c|}{2} & \multicolumn{2}{|c|}{4} & \multicolumn{2}{|c|}{8} & \multicolumn{2}{|c|}{16} & \multicolumn{2}{|c|}{32} & \multicolumn{2}{|c|}{64}\\

        \hline 
         add20       & \numprint{621}            & \textbf{\numprint{586}}   & \textbf{\numprint{1158}}  & \numprint{1159}            & \textbf{\numprint{1695}}  & \numprint{1696}            & +\textbf{\numprint{2057}} & \numprint{2062}            & \textbf{\numprint{2399}}    & \numprint{2687}           & \textbf{\numprint{3004}}   & \numprint{3108} \\
 data        & \textbf{\numprint{188}}   & \numprint{188}            & \textbf{\numprint{376}}   & \numprint{377}             & \textbf{\numprint{656}}   & \numprint{656}             & \textbf{\numprint{1124}}  & \numprint{1135}            & \textbf{\numprint{1812}}    & \numprint{1858}           & \textbf{\numprint{2863}}   & \numprint{2885} \\
 3elt        & \textbf{\numprint{89}}    & \numprint{89}             & \textbf{\numprint{199}}   & \numprint{199}             & \textbf{\numprint{340}}   & \numprint{340}             & \textbf{\numprint{568}}   & \numprint{568}             & \textbf{\numprint{956}}     & \numprint{967}            & \textbf{\numprint{1539}}   & \numprint{1553} \\
 uk          & \textbf{\numprint{19}}    & \numprint{19}             & \textbf{\numprint{40}}    & \numprint{40}              & \numprint{81}             & @\textbf{\numprint{80}}    & \textbf{\numprint{144}}   & \numprint{144}             & \textbf{\numprint{249}}     & \numprint{251}            & \textbf{\numprint{410}}    & \numprint{417} \\
 add32       & \textbf{\numprint{10}}    & \numprint{10}             & \textbf{\numprint{33}}    & \numprint{33}              & \textbf{\numprint{66}}    & \numprint{66}              & \textbf{\numprint{117}}   & \numprint{117}             & \textbf{\numprint{212}}     & \numprint{212}            & \numprint{487}             & @\textbf{\numprint{486}} \\
 bcsstk33    & \textbf{\numprint{10097}} & \numprint{10097}          & \textbf{\numprint{21338}} & \numprint{21508}           & \textbf{\numprint{34175}} & \numprint{34178}           & \textbf{\numprint{54696}} & \numprint{54763}           & \textbf{\numprint{77682}}   & \numprint{77964}          & \textbf{\numprint{108261}} & \numprint{108467} \\
 whitaker3   & \textbf{\numprint{126}}   & \numprint{126}            & \textbf{\numprint{380}}   & \numprint{380}             & \textbf{\numprint{654}}   & \numprint{654}             & \textbf{\numprint{1086}}  & \numprint{1091}            & \textbf{\numprint{1667}}    & \numprint{1678}           & \textbf{\numprint{2507}}   & \numprint{2532} \\
 crack       & \textbf{\numprint{183}}   & \numprint{183}            & \textbf{\numprint{362}}   & \numprint{362}             & \textbf{\numprint{676}}   & \numprint{676}             & \textbf{\numprint{1086}}  & \numprint{1089}            & \textbf{\numprint{1670}}    & \numprint{1687}           & \textbf{\numprint{2536}}   & \numprint{2555} \\
 wing\_nodal & \textbf{\numprint{1695}}  & \numprint{1695}           & \textbf{\numprint{3563}}  & \numprint{3563}            & \textbf{\numprint{5401}}  & \numprint{5422}            & \textbf{\numprint{8307}}  & \numprint{8339}            & +\textbf{\numprint{11818}}  & \numprint{11828}          & \textbf{\numprint{15996}}  & \numprint{16124} \\
 fe\_4elt2   & \textbf{\numprint{130}}   & \numprint{130}            & \textbf{\numprint{349}}   & \numprint{349}             & \textbf{\numprint{603}}   & \numprint{603}             & \textbf{\numprint{1000}}  & \numprint{1002}            & \textbf{\numprint{1608}}    & \numprint{1620}           & \textbf{\numprint{2487}}   & \numprint{2519} \\
 vibrobox    & \numprint{11538}          & \textbf{\numprint{10310}} & \numprint{18967}          & @\textbf{\numprint{18956}} & \numprint{24555}          & @\textbf{\numprint{24422}} & \numprint{32867}          & @\textbf{\numprint{32102}} & \numprint{41393}            & *\textbf{\numprint{40085}} & \numprint{47876}           & *\textbf{\numprint{47651}} \\
 bcsstk29    & \textbf{\numprint{2818}}  & \numprint{2818}           & \textbf{\numprint{8029}}  & \numprint{8029}            & \numprint{14113}          & @\textbf{\numprint{13904}} & \numprint{22126}          & \textbf{\numprint{21768}}  & \textbf{\numprint{34817}}   & \numprint{34841}          & \textbf{\numprint{55887}}  & \numprint{57031} \\
 4elt        & \textbf{\numprint{138}}   & \numprint{138}            & \textbf{\numprint{320}}   & \numprint{320}             & \textbf{\numprint{532}}   & \numprint{532}             & \textbf{\numprint{929}}   & \numprint{932}             & \textbf{\numprint{1544}}    & \numprint{1547}           & \textbf{\numprint{2559}}   & \numprint{2574} \\
 fe\_sphere  & \textbf{\numprint{386}}   & \numprint{386}            & \textbf{\numprint{766}}   & \numprint{766}             & \textbf{\numprint{1152}}  & \numprint{1152}            & \textbf{\numprint{1709}}  & \numprint{1709}            & \textbf{\numprint{2482}}    & \numprint{2488}           & \textbf{\numprint{3560}}   & \numprint{3584} \\
 cti         & \textbf{\numprint{318}}   & \numprint{318}            & \textbf{\numprint{944}}   & \numprint{944}             & \textbf{\numprint{1746}}  & \numprint{1749}            & \textbf{\numprint{2788}}  & \numprint{2837}            & \textbf{\numprint{4022}}    & \numprint{4117}           & \textbf{\numprint{5695}}   & \numprint{5818} \\
 memplus     & \textbf{\numprint{5478}}  & \numprint{5484}           & \numprint{9597}           & @\textbf{\numprint{9448}}  & \numprint{11994}          & @\textbf{\numprint{11776}} & \numprint{13107}          & *\textbf{\numprint{13001}}  & \numprint{14436}            & *\textbf{\numprint{14107}} & \textbf{\numprint{16269}}  & \numprint{16543} \\
 cs4         & \textbf{\numprint{366}}   & \numprint{366}            & \numprint{928}            & @\textbf{\numprint{925}}   & \textbf{\numprint{1434}}  & \numprint{1436}            & \numprint{2088}           & @\textbf{\numprint{2087}}  & \textbf{\numprint{2903}}    & \numprint{2910}           & \textbf{\numprint{3982}}   & \numprint{4032} \\
 bcsstk30    & \textbf{\numprint{6335}}  & \numprint{6335}           & \textbf{\numprint{16583}} & \numprint{16596}           & \textbf{\numprint{34566}} & \numprint{34577}           & \textbf{\numprint{69915}} & \numprint{70604}           & +\textbf{\numprint{113742}} & \numprint{113788}         & \textbf{\numprint{172654}} & \numprint{172929} \\
 bcsstk31    & \textbf{\numprint{2699}}  & \numprint{2699}           & \numprint{7283}           & @\textbf{\numprint{7282}}  & \textbf{\numprint{13178}} & \numprint{13201}           & \textbf{\numprint{23628}} & \numprint{23761}           & \textbf{\numprint{37408}}   & \numprint{37652}          & \numprint{58085}           & *\textbf{\numprint{58076}} \\
 fe\_pwt     & \textbf{\numprint{340}}   & \numprint{340}            & \textbf{\numprint{704}}   & \numprint{704}             & \textbf{\numprint{1432}}  & \numprint{1433}            & \textbf{\numprint{2797}}  & \numprint{2797}            & \textbf{\numprint{5514}}    & \numprint{5523}           & \textbf{\numprint{8152}}   & \numprint{8222} \\
 bcsstk32    & \textbf{\numprint{4667}}  & \numprint{4667}           & \numprint{9208}           & @\textbf{\numprint{9195}}  & \textbf{\numprint{20200}} & \numprint{20204}           & \textbf{\numprint{35619}} & \numprint{35936}           & \textbf{\numprint{60127}}   & \numprint{60776}          & \textbf{\numprint{91361}}  & \numprint{91863} \\
 fe\_body    & \textbf{\numprint{262}}   & \numprint{262}            & \textbf{\numprint{598}}   & \numprint{598}             & \textbf{\numprint{1023}}  & \numprint{1026}            & \numprint{1736}           & @\textbf{\numprint{1714}}  & \textbf{\numprint{2788}}    & \numprint{2935}           & \textbf{\numprint{4723}}   & \numprint{4825} \\
 t60k        & \textbf{\numprint{75}}    & \numprint{75}             & \textbf{\numprint{208}}   & \numprint{208}             & \textbf{\numprint{454}}   & \numprint{454}             & \textbf{\numprint{805}}   & \numprint{805}             & \textbf{\numprint{1313}}    & \numprint{1320}           & \textbf{\numprint{2062}}   & \numprint{2079} \\
 wing        & \textbf{\numprint{784}}   & \numprint{784}            & \numprint{1634}           & @\textbf{\numprint{1610}}  & \textbf{\numprint{2474}}  & \numprint{2479}            & \numprint{3860}           & @\textbf{\numprint{3857}}  & \numprint{5606}             & @\textbf{\numprint{5584}}  & \textbf{\numprint{7643}}   & \numprint{7656} \\
 brack2      & \textbf{\numprint{708}}   & \numprint{708}            & \textbf{\numprint{3013}}  & \numprint{3013}            & \textbf{\numprint{7043}}  & \numprint{7099}            & \textbf{\numprint{11576}} & \numprint{11636}           & \textbf{\numprint{17323}}   & \numprint{17398}          & \textbf{\numprint{25911}}  & \numprint{25913} \\
 finan512    & \textbf{\numprint{162}}   & \numprint{162}            & \textbf{\numprint{324}}   & \numprint{324}             & \textbf{\numprint{648}}   & \numprint{648}             & \textbf{\numprint{1296}}  & \numprint{1296}            & \textbf{\numprint{2592}}    & \numprint{2592}           & \textbf{\numprint{10560}}  & \numprint{10560} \\
 fe\_tooth   & \textbf{\numprint{3814}}  & \numprint{3814}           & \numprint{6848}           & @\textbf{\numprint{6846}}  & \textbf{\numprint{11372}} & \numprint{11408}           & \textbf{\numprint{17265}} & \numprint{17396}           & \numprint{25096}            & *\textbf{\numprint{24933}} & \numprint{34625}           & *\textbf{\numprint{34433}} \\
 fe\_rotor   & \textbf{\numprint{2031}}  & \numprint{2031}           & \textbf{\numprint{7171}}  & \numprint{7180}            & \numprint{12804}          & @\textbf{\numprint{12726}} & \textbf{\numprint{20259}} & \numprint{20438}           & \textbf{\numprint{30995}}   & \numprint{31233}          & \numprint{45948}           & *\textbf{\numprint{45911}} \\
 598a        & \textbf{\numprint{2388}}  & \numprint{2388}           & \numprint{7957}           & @\textbf{\numprint{7948}}  & \numprint{15935}          & *\textbf{\numprint{15924}}  & \textbf{\numprint{25624}} & \numprint{25741}           & \numprint{39225}            & *\textbf{\numprint{38627}} & \numprint{56787}           & *\textbf{\numprint{56179}} \\
 fe\_ocean   & \textbf{\numprint{387}}   & \numprint{387}            & \textbf{\numprint{1813}}  & \numprint{1816}            & \textbf{\numprint{4063}}  & \numprint{4091}            & \numprint{7849}           & *\textbf{\numprint{7771}}   & \textbf{\numprint{12597}}   & \numprint{12711}          & \numprint{19992}           & *\textbf{\numprint{19989}} \\
 144         & \textbf{\numprint{6478}}  & \numprint{6478}           & \numprint{15156}          & @\textbf{\numprint{15140}} & \textbf{\numprint{25240}} & \numprint{25273}           & \textbf{\numprint{37750}} & \numprint{37896}           & \textbf{\numprint{56059}}   & \numprint{56550}          & \textbf{\numprint{78538}}  & \numprint{79198} \\
 wave        & \textbf{\numprint{8657}}  & \numprint{8658}           & \textbf{\numprint{16747}} & \numprint{16780}           & \textbf{\numprint{28758}} & \numprint{28979}           & \textbf{\numprint{42354}} & \numprint{42516}           & \numprint{61132}            & @\textbf{\numprint{61104}} & \textbf{\numprint{84353}}  & \numprint{85589} \\
 m14b        & \textbf{\numprint{3826}}  & \numprint{3826}           & \numprint{12988}          & @\textbf{\numprint{12973}} & \textbf{\numprint{25666}} & \numprint{25690}           & \textbf{\numprint{42295}} & \numprint{42351}           & \numprint{65852}            & @\textbf{\numprint{65835}} & \textbf{\numprint{97550}}  & \numprint{98211} \\
 auto        & \numprint{9985}           & @\textbf{\numprint{9949}}  & \numprint{26691}          & @\textbf{\numprint{26614}} & \textbf{\numprint{45442}} & \numprint{45470}           & \textbf{\numprint{76542}} & \numprint{77005}           & \textbf{\numprint{120490}}  & \numprint{121032}         & \textbf{\numprint{171880}} & \numprint{172167} \\

        \hline
        \end{tabular}
        \end{center} \caption{Computing partitions from scratch $\epsilon = 1$\%. In each $k$-column the results computed by KaBaPE are on the left and the current Walshaw cuts are presented on the right side. Entries marked with a * can be improved by our refinement techniques when using the current record as input. Entries marked with a + have been obtained using different parameters and seeds. @ indicates that we hold this record.}

        \end{table}
        \end{landscape}
        %\vfill
        %\vspace*{5cm}

\end{appendix}
\end{document}